\documentclass[prd,twocolumn,superscriptaddress,preprintnumbers,nofootinbib]{revtex4-1}

\pdfoutput=1

\usepackage{amsfonts,amssymb,amsmath}
\usepackage{bm}
\usepackage{bbm}
\usepackage{epsfig}

\newcommand{\avg}[1]{\langle#1\rangle}

\newcommand{\wba}[1]{{\overline{#1}}}

\newcommand{\cO}{\mathcal{O}}

\newcommand{\U}{{\rm U}}
\newcommand{\SU}{{\rm SU}}
\newcommand{\SUC}{{\rm SU(3)_C}}
\newcommand{\SUL}{{\rm SU(2)_L}}
\newcommand{\UY}{{\rm U(1)_Y}}

\newcommand{\cc}{\text{c.c.}}

\newcommand{\UR}{{\rm U(1)_R}}
\newcommand{\UX}{{\rm U(1)_X}}

\newcommand{\MP}{M_\text{Pl}}

\newcommand{\ra}{\rightarrow}
\newcommand{\lsim}{\lesssim}

\newcommand{\tev}{~\text{TeV}}
\newcommand{\gev}{~\text{GeV}}

\newcommand{\Rch}[1]{\textbf{R}[#1]}

\newcommand{ \slashchar }[1]{\setbox0=\hbox{$#1$}   
   \dimen0=\wd0                                     
   \setbox1=\hbox{/} \dimen1=\wd1                   
   \ifdim\dimen0>\dimen1                            
      \rlap{\hbox to \dimen0{\hfil/\hfil}}          
      #1                                            
   \else                                            
      \rlap{\hbox to \dimen1{\hfil$#1$\hfil}}       
      /                                             
   \fi}                                             %

\DeclareMathOperator{\casim}{C_2^a}

\begin{document}

\title{Viable Gravity-Mediated Supersymmetry Breaking}

\author{Graham D. Kribs}
\affiliation{Department of Physics, University of Oregon, Eugene, OR
  97403} 
\author{Takemichi Okui}
\affiliation{Department of Physics, Florida State University,
  Tallahassee, FL 32306} 
\author{Tuhin S. Roy}
\affiliation{Department of Physics, University of Oregon, Eugene, OR
  97403} 

\begin{abstract}

We present a complete, viable model of gravity-mediated supersymmetry 
breaking that is safe from all flavor constraints.
The central new idea is to employ a supersymmetry breaking sector
without singlets, but with $D$-terms comparable to $F$-terms, 
causing supersymmetry breaking to be dominantly communicated 
through $\UR$-symmetric operators.  We construct a visible sector
that is an extension of the MSSM where an \emph{accidental}
$\UR$-symmetry emerges naturally.  
Gauginos acquire Dirac masses from gravity-mediated $D$-terms, 
and tiny Majorana masses from anomaly-mediated contributions.  
Contributions to soft breaking scalar
$({\rm mass})^2$ arise from flavor-arbitrary gravity-induced $F$-terms
plus one-loop finite flavor-blind contributions from Dirac gaugino
masses.  Renormalization group evolution of the gluino causes it
to naturally increase nearly an order of magnitude larger than
the squark masses.  This hierarchy, combined with an accidentially 
$\UR$-symmetric visible sector, nearly eliminates all flavor violation
constraints on the model.  If we also freely tune couplings and phases
within the modest range $0.1$-$1$, while maintaining nearly 
flavor-anarchic Planck-suppression contributions, we find our model to be
safe from $\Delta m_K$, $\epsilon_K$, and $\mu \ra e$ lepton flavor 
violation.  Dangerous $\UR$-violating K\"ahler operators in the Higgs sector
are eliminated through a new gauged $\UX$ symmetry that is
spontaneously broken with electroweak symmetry breaking.  
Kinetic mixing between $\UX$ and $\UY$ is present with
loop-suppressed (but log-enhanced) size $\epsilon$.  The 
$Z'$ associated with this $\UX$ has very peculiar couplings
-- it has order one strength to Higgs doublets and approximately
$\epsilon$ strength to hypercharge.  The $Z'$ could be remarkably light
and yet have escaped direct and indirect detection.

\end{abstract}

\maketitle

\section{Introduction}

If supersymmetry plays a role in the physics near the electroweak scale, 
the most pressing question is how supersymmetry breaking is mediated
to the ``visible'' supersymmetrized standard model.
The nature of this mediation makes a qualitative impact on the 
superpartner spectrum and interactions, thereby being directly relevant 
to the physics probed by experiments including the Large Hadron Collider
(LHC).  

Gravity
mediation~\cite{Chamseddine:1982jx,Barbieri:1982eh,Ibanez:1982ee,Hall:1983iz,Ohta:1982wn,Ellis:1982wr,AlvarezGaume:1983gj,Nilles:1983ge,Nath:1983fp}
is far and away the simplest 
mediation mechanism.  The hidden and visible sectors interact only 
through Planck-suppressed nonrenormalizable operators.   
Gaugino and scalar superpartners acquire comparable masses. 
Moreover, unlike low scale mediation, the $\mu$ and $B_\mu$
terms in the Higgsino and Higgs sector can easily acquire
comparable sizes~\cite{Giudice:1988yz}.  
Also, with the gravitino comparable to other soft supersymmetry breaking 
masses, a neutral superpartner of the visible sector can be the 
lightest supersymmetric particle (LSP), serving as a dark matter candidate. 
Last, but not least, gauge coupling 
unification~\cite{Dimopoulos:1981zb,Dimopoulos:1981yj} is automatic.   

Gravity mediation does, however, have a serious flavor problem.  
Since the mediation of supersymmetry breaking is at the Planck scale,
this is necessarily above (or at) the scale of generating flavor.
No symmetries enforce flavor universality in the operators mediating
supersymmetry breaking.  Thus, there is no mechanism to ensure
the mass-squared matrices of squarks and sleptons 
to be almost exactly proportional to an
identity matrix for consistency with constraints
on flavor violation beyond the standard model%
~\cite{Gabbiani:1988rb,Gabbiani:1996hi,Bagger:1997gg,Ciuchini:1998ix}.  
Even though models with continuous flavor symmetries are proposed to
avoid the flavor problem~\cite{Barbieri:1995uv,Barbieri:1996ww}, it is
unlikely that gravitational interactions respect any global
symmetry~\cite{Kallosh:1995hi}. 
Models in which squark and slepton squared-mass matrices are aligned
with the corresponding Yukawa matrices~\cite{Hall:1990ac,Nir:1993mx} 
because of gauged discrete family
symmetries~\cite{Kaplan:1993ej,ArkaniHamed:1996xm} do solve the flavor
problem but these implementations are rather complicated.

In this paper we present a complete model of supersymmetry breaking
with gravity mediation and a visible sector that is safe from flavor 
constraints.
We exploit the observation that flavor violation beyond the
standard model can be greatly suppressed 
\emph{even without flavor universality of the squark and slepton sectors} 
if the visible sector particle content respects an approximate global $\UR$
symmetry~\cite{Kribs:2007ac}.  
For us, the origin of the approximate $\UR$ symmetry is \emph{accidental}.
We do not enforce $\UR$ symmetry per se, but arrange the
hidden and visible sectors such that the only potentially dangerous
$\UR$-violation arises as a result of anomaly mediation 
(which, nevertheless, is small enough to maintain flavor safety).
Getting this to work with gravity mediation -- namely,
sizeable $D$-terms, $F$-terms, a modest hierarchy between the
Dirac gluino and squark masses, not generating sizeable Majorana masses,
but generating approximately $R$-symmetric Higgsino masses
is the subject of the paper.  Our philosophy is to exploit
holomorphy of the superpotential (any terms we do not need or want
can be naturally set to zero), but to strictly eliminate dangerous
K\"ahler terms through symmetries.  As a consequence, 
we are inevitably led to gauging an additional $\UX$ symmetry 
in the visible sector.  The phenomenology of this $\UX$ is 
quite interesting:  the $Z'$ of this broken $\UX$ couples with 
order one strength to the Higgs sector and order $10^{-4} \ra 10^{-2}$ 
strength to standard model fields. 

Our paper is organized as follows.  In Sec.~\ref{philosophy-sec},
we outline the basic philosophy regarding the hidden sector,
the visible sector, the mediation, the sizes of dimensionless
constants, and so forth.  In Sec.~\ref{susy-breaking-sec} 
we present the supersymmetry breaking sector -- the ``$4$-$1$'' model -- 
and its relevant characteristics to our framework. 
In Sec.~\ref{mediation-sec} we discuss the full set of operators
that can mediate supersymmetry breaking up to dimension-6.
This provides the basis to understand the content and 
supersymmetry breaking induced into the visible sector, 
discussed in Sec.~\ref{visible-sec}.  It is here where we show
how the visible sector can remain approximately $R$-symmetric
despite the spontaneous (and explicit) $R$-violation of the
hidden sector.  We then outline the main phenomenological 
implications of our model in Sec.~\ref{sec:pheno}.  The 
gaugino and scalar spectrum are discussed in the context of a
benchmark point, illustrating that all flavor constraints 
can be satisfied.  Discussion of the Higgs sector, as well as
the $\UX$ sector phenomenology is begun.  Finally, we present
a brief discussion of unification in Sec.~\ref{unifon-sec} and conclude
with a more general discussion in Sec.~\ref{discussion-sec}.

\section{Philosophy}
\label{philosophy-sec}

The basic philosophy behind our model-building efforts can be summarized
as follows:
\begin{itemize}
\item The supersymmetry breaking hidden sector is specified:
  we employ the $\SU(4) \times \U(1)$ (``$4$-$1$'')
  model~\cite{Dine:1995ag}. 
  This model has two virtues for our proposal:  First, it 
  contains no singlet fields.  Second, it contains a 
  $\U(1)$ that acquires a $D$-term comparable to the $F$-terms. 
\item Supersymmetry breaking is only mediated by Planck-suppressed operators
  involving $4$-$1$ fields and visible sector fields. 
\item ``Chiral'' K\"ahler potential%
\footnote{By ``chiral'' K\"ahler we refer to the operators that appear 
as $\int\!\!d^4 \theta \ f$ in the global supersymmetric limit, 
where $f$ are polynomials of only chiral or
antichiral superfields.} 
  and all superpotential operators
  are present or absent as we deem appropriate for the model.  
  We exploit the technical naturalness of these operators to 
  eliminate some otherwise potentially dangerous mediation operators.  
\item All ``nonchiral'' operators in the K\"ahler potential allowed
  by symmetries are assumed to be present. 
\item Nonzero Planck-suppressed operators, and dimensionless couplings, 
  have coefficients between 
  about $0.1$ to $1$.  We allow ourselves complete freedom in
  setting the sizes of different operators with coefficients in
  this range.
\end{itemize}

The $4$-$1$ model is one of several examples of global supersymmetry 
breaking models that have a $D$-term generated with size of order 
an $F$-term~\cite{Dine:1995ag,Carpenter:2005tz,Dumitrescu:2010ca}.
For us, the crucial feature of this model
is the absence of hidden sector singlets.  
Generically, at the local supersymmetry breaking minimum, the hidden sector
fields acquire scalar and supersymmetry breaking expectation values
that spontaneously break the $\UR$ symmetry, in accord with
Ref.~\cite{Nelson:1993nf}.  Operators mediating 
supersymmetry breaking involving supersymmetry breaking spurions 
include one or more powers of the $D$-term, which automatically 
preserves $\UR$, and $F$-terms, which must arise in gauge-invariant
combinations.  For us, the absence of singlets 
implies the lowest dimension gauge-invariants are
also $\UR$-invariant (of the form $X^\dagger X$, where $X$ is
a hidden sector chiral superfield, as we will see).  Hence, despite
the sizeable  spontaneous breaking of $\UR$ in the hidden sector, the
lowest dimension operators mediating supersymmetry breaking do so
through $\UR$-symmetric combinations of hidden sector 
fields.  This observation was also exploited in a 
different context in Ref.~\cite{ArkaniHamed:2004yi}.

Embedding a global supersymmetry breaking model into supergravity
also requires adding a constant term to the superpotential, 
explicitly breaking $\UR$, to fine-tune the cosmological constant
to zero.  The explicit breaking gives a mass to the would-be
massless $R$-axion~\cite{Bagger:1994hh}.
It also leads to the gravitino
acquiring a Majorana mass, absorbing the would-be massless goldstino
from supersymmetry breaking.  The visible sector is thus not 
completely isolated from explicit $\UR$ breaking, due to the anomaly-mediated 
contribution to visible sector fields proportional to the 
gravitino mass~\cite{Giudice:1998xp,Randall:1998uk}.

One might think that the approach should be to simply demand an 
approximate $\UR$ symmetry on the visible sector by hand.  But one could 
also hold the same viewpoint for flavor universality; 
if we demand the $\U(3)^5$ flavor symmetry on the
soft masses in isolation, flavor universality would be approximately
maintained even though the Yukawa couplings violate
$\U(3)^5$. Therefore, if we wish to solve the flavor problem without
appeal to an ``initial condition'' at the Planck scale, we must
seek a mechanism where an approximate $\UR$ symmetry in the visible
sector emerges.  Only then can we compare the
$\UR$-symmetric solution to the flavor problem with {\it e.g.} gauge 
mediation~\cite{Dine:1981za,Dine:1981gu,Dimopoulos:1980yf,Nappi:1982hm,AlvarezGaume:1981wy,Dimopoulos:1982gm,Dine:1993yw,Dine:1994vc,Dine:1995ag}, gaugino 
mediation~\cite{Kaplan:1999ac,Chacko:1999mi} or anomaly  
mediation~\cite{Giudice:1998xp,Randall:1998uk},
in which the $\U(3)^5$-preserving soft masses emerge 
in the presence of the $\U(3)^5$-violating Yukawa couplings.

\section{Supersymmetry Breaking}
\label{susy-breaking-sec}

We employ the $4$-$1$~model for our supersymmetry
breaking sector. The $4$-$1$ model is an $\SU(4)\otimes\U(1)$
supersymmetric gauge theory with four chiral superfields transforming
under $\SU(4)\otimes\U(1)$ as  
\begin{equation}
  \label{eq:hidden-mult}
  \centering
  \begin{tabular}{ c  || c | c }
      \ fields \ & \ SU$(4)$ \  & \ U$(1)$  \\
      \hline
        $X_1$ \ & $4$ & $-3$ \\
        $X_2$ \ &  $\bar 4$ & $-1$ \\
        $X_3$ \ & $6$ & $2$ \\
        $X_4$ \ & $1$ & $4$ \\
       \hline 
  \end{tabular}
\end{equation}
At high energies, the theory is weakly coupled and the dimensions of 
all the chiral superfields are close to their classical dimensions.  
The field content allows a unique renormalizable superpotential 
$\lambda X_1 X_2 X_4$ permitted under the symmetries. 
At lower energies, $\SU(4)$ gets strong at a scale $\Lambda_4$.
The dynamically generated non-perturbative superpotential
consistent with all the symmetries is unique,
\begin{equation}
  \label{eq:w-np}
  W_{\text{np}} = \frac{\Lambda_4^5}{\sqrt{X_1 X_2 X_3^2}}
\end{equation}

In the limit $g_4 \gg \lambda \gg g_1$, the minimum of the potential
occurs in the $D$-flat direction of $\SU(4)$. By a suitable gauge
rotation, the non-zero vevs have the form 
\begin{equation}
  \label{eq:hidden-vev}
  X_1 =  X_2^\text{T} =
     \begin{bmatrix} v_1 \\ 0 \\ 0 \\ 0 \end{bmatrix}  \quad 
   X_3 =  \begin{bmatrix} v_2 \sigma_2 & 0 \\
                                  0  & v_2 \sigma_2
                                  \end{bmatrix} \quad 
    X_4 = v_3                                
\end{equation}
After rescaling all the fields, $\phi \rightarrow
\frac{\Lambda_4}{\lambda^{1/4}} \phi$, the scalar potential
is~\cite{Dine:1995ag,Carpenter:2005tz}
\begin{multline}
 V = V_F + V_D  \text{ where } V_F = \sum_{i} \left| F_{X_i} \right|^2 
   \text{ and }  V_D = \frac{1}{2}D_1^2 \\
  V_F = \lambda^{6/5} \Lambda_4^4 \ \Bigg[ 
     \left| v_1 \right|^2 +\left| 2 v_1 v_3 - \frac{1}{v_1^2 v_2}
        \right|^2 + \left| \frac{1}{v_1 v_2^2} \right|^2 \Bigg] \\
  V_D =  \frac{1}{2}  \ \Bigg[  4 g_1 \ \frac{
      \Lambda_4^2}{\lambda^{2/5}}\ 
    \Big(  \left| v_1 \right|^2 - \left| v_2\right|^2 - 
        \left| v_3 \right|^2 \Big) \Bigg]^2
\end{multline}
Both $V_F$ and $V_D$ cannot be simultaneously zero for any value of
${v_1,v_2,v_3}$ and consequently supersymmetry is broken at the minimum. 
We can find the minimum of the potential numerically:  
For $g_1/\lambda \lesssim 0.1$, the potential is minimized at $\avg{D_1}
\sim 1.5 \: \text{Max}[\avg{F_{X_i}}] \neq 0$. 

Therefore, the $4$-$1$ model provides both $F$-term and $D$-term 
supersymmetry breaking.  Importantly, the absence of any gauge-singlet 
$F$-term vevs in the $4$-$1$ model implies restrictions on the
form of the operators that can mediate supersymmetry breaking.
Amusingly, in the original context of gauge mediation, the $4$-$1$
model was abandoned precisely because of the absence of singlets,
since no Majorana gaugino masses are generated.
This ``bug'' for gauge-mediation becomes an essential ``feature'' 
of our model that allows us to realize an accidental $\UR$ symmetry.

\section{Mediating Supersymmetry Breaking}
\label{mediation-sec}

Supersymmetry breaking is mediated through 
Planck-suppressed operators involving the visible sector fields.  
In this section we list and categorize all possible contact operators 
suppressed by up to two powers of $\MP$.     

Below, $X$ represents any one of chiral superfields
$X_1,X_2,X_3,X_4$ in the $4$-$1$ model. 
As shown in Eq.~\eqref{eq:hidden-mult}, $X$ is 
\emph{not} a singlet and so all operators involving single power of
$X$ are forbidden. Majorana gaugino mass operators 
(namely, $\int\!\! d^2 \theta X \, W_a W_a$), 
the $A~$term operators ($\int\!\! d^2 \theta X \, QUH$) 
and the Giudice-Masiero $\mu~$term ($\int\!\! d^4 \theta X^\dag \, H_u
H_d$)  involve a single $X$ and hence do not exist in our model. 
The absence of these operators will be essential for the
emergence of an accidental $\UR$.
The gauge-invariant chiral combination 
$X_1 X_2 X_4$ can give rise to these operators but they are suppressed 
by additional powers of $\MP$ and are inconsequential.  

The leading gauge-invariant operator involving $X$ is $X^\dag X$,
by which we refer to the operator 
$X_i^\dag \, \exp \left( q_a^i g_a V_a \right) \, X_i$, where $q_a^i$
is the charge of $X_i$ under the ``$a$''th gauge group.
This combination is also $\UR$-invariant. 

If the visible sector is extended to include chiral superfields
$\Sigma_a$ in the adjoint representation of the gauge group ``$a$'',
gauginos can acquire $\UR$-invariant Dirac masses through 
dimension~$5$ operators involving the hidden sector $\U(1)$ 
superfield strength
$W'$~\cite{Polchinski:1982an,Hall:1990hq,Randall:1992cq,Dine:1992yw,Fox:2002bu}.
\begin{equation}
  \label{eq:dirac}
  \int\!\! d^2\!\theta \ \frac{W'}{\MP} W_a 
                   \Sigma_a \; . 
\end{equation}

For the rest of this section, we use $\Phi_i$ to refer to all visible
sector chiral superfields ({\it i.e.} the index $i$ runs over all
matter, Higgs and $\Sigma$ fields).  
All dimension~$5$ and $6$  operators in the superpotential and
K\"ahler potential 
involving fields from both the hidden and visible sectors 
can be written as: 
\begin{align}
\text{dim 5:}   \qquad    
  & \int\!\! d^2\!\theta \ \frac{W' W'}{\MP} \Phi_i 
    \label{op2} \\
  & \int\!\! d^2\!\theta \ \frac{X_1 X_2 X_4}{\MP} \Phi_i 
   \label{op3} \\ 
  & \int\!\! d^4\!\theta \ \frac{X^\dag X}{\MP} \Phi_i 
   \label{op4} \\
\text{dim 6:}   \qquad 
  & \int\!\! d^2\!\theta \ \frac{W' W'}{\MP^2} \Phi_i \Phi_j 
   \label{op5} \\
  & \int\!\! d^2\!\theta \ \frac{X_1 X_2 X_4}{\MP^2} \Phi_i \Phi_j
   \label{op6} \\ 
  & \int\!\! d^4\!\theta \ \frac{X^\dag X}{\MP^2} \Phi_i \Phi_j 
   \label{eq:b} \\
  & \int\!\! d^4\!\theta \ \frac{X^\dag X}{\MP^2} \Phi_i^\dag \Phi_j 
   \label{eq:scalar} 
\end{align}
All the dimension$~5$ operators in Eqs.~(\ref{op2})-(\ref{op4}) 
involve only the gauge singlets of $\Phi$ (for example, $\Sigma_1$).  
Operators in  Eqs.~(\ref{op5})-(\ref{eq:b}) 
involve either two chiral fields of opposite charge (for example, $H_u
H_d$) or chiral fields in the real representation ($\Sigma_a^2$). In
particular, the $B_\mu$ term (namely, $  \int\!\! 
d^4\theta X^\dag X \, H_u H_d $)  is contained in Eq.~\eqref{eq:b}. 
Finally, Eq.~\eqref{eq:scalar} involves every field of $\Phi$ and
generates soft mass-squareds for all the scalars.

\section{The Visible Sector}
\label{visible-sec}

Having established that the visible sector is now shielded from
$\UR$ breaking in the hidden sector, if we construct
the visible sector to be $\UR$-invariant itself, then it 
will remain so after supersymmetry breaking.  

We describe our $\UR$-invariant visible sector in terms of its 
three subsectors:
the gauge and gaugino sector, the matter (squark and slepton) sector, 
and the Higgs sector.  In this section we first concentrate on the 
supersymmetry breaking contributions from gravity-mediated operators, 
and only later reintroduce the small corrections arising from 
anomaly-mediation.  

The contact operators suppressed by $\MP$ must be renormalized down
to the electroweak symmetry breaking (EWSB) scale before they are 
converted to soft mass terms and the
flavor violating observables are calculated. In our model, we need to 
renormalize the Dirac gauginos mass operators in Eq.~\eqref{eq:dirac}, 
scalar soft mass squared operators in Eq.~\eqref{eq:scalar} 
and the B-type scalar mass operators in Eq.~\eqref{eq:b}. 
As we will see, the renormalization of these operators leads to a 
substantial increase in the gluino Dirac mass relative to the squark
masses even through $\avg{D} \sim \avg{F}$ at the Planck scale.  
Although these operators scale due to the interactions
present in the visible sector as well as in the hidden
sector~\cite{Cohen:2006qc}, we do not consider renormalization 
from hidden sector interactions since the effect is small in 
the $4$-$1$ model. 
Amusingly, however, the effect
of hidden sector renormalization is to further lower the 
contributions to the soft mass squared of the matter (and other scalar)
sector relative to the Dirac gauginos, allowing for potentially
an even larger mass ratio between gauginos and squarks/sleptons%
~\cite{Luty:2001jh,Luty:2001zv,Dine:2004dv,Ibe:2005pj,Ibe:2005qv,Schmaltz:2006qs}.

\subsection{Gauge and Gaugino Sector}

Since Majorana masses are not generated by gravity-mediation from
the $4$-$1$ model, we introduce chiral adjoint superfields
$\Sigma_a$ for each gauge symmetry of the standard model%
~\cite{Polchinski:1982an,Hall:1990hq,Randall:1992cq,Fox:2002bu,Nelson:2002ca,Chacko:2004mi,Carone:2005iq,Nomura:2005rj,Nomura:2005qg,Carpenter:2005tz,Antoniadis:2006uj,Nakayama:2007cf,Amigo:2008rc}. 
This allows us 
to write the gravity-mediated operator Eq.~\eqref{eq:dirac},
which generates Dirac gaugino masses (as well as scalar masses
for the adjoint scalar). 

The part of the Lagrangian involving visible sector 
gauge field-strength superfields $W_a$ is then written as 
\begin{equation}
   \label{eq:int-gauge}
    \int \!\! d^2 \! \theta \ \sum_a   W^a W^a \ +\
      \int\!\! d^2\!\theta \ \sum_a \sqrt{2} \: w_a \frac{W'}{\MP} W_a
                   \Sigma_a \, , 
\end{equation}
where $w_a$ are order one dimensionless coupling constants. These
constants  -- leading to the Dirac masses -- 
scale differently from operators
that would lead to Majorana masses in models with gauge singlets
in the hidden sector~\cite{Fox:2002bu}.
The evolution is given by
\begin{multline}
  \label{eq:ren-dirac}
  w_a \left(\mu \right) = w_a \left(\MP
  \right) \sqrt{ \frac{Z_{\Sigma_a}\left(\MP \right)} 
                      {Z_{\Sigma_a}\left(\mu \right)} }
          \sqrt{ \frac{\alpha_a\left( \MP \right)} 
                      {\alpha_a\left( \mu \right)} } \\
               =       w_a \left(\MP
  \right) \begin{cases} 
\left(\frac{\mu}{\Lambda}\right)^{-\frac{N_a}{2 \pi}\alpha_a} &
\text{for }b_a = 0, \\
\left(\frac{g_a(\mu)}{g_a(\Lambda)}\right)^{1-\frac{2N_a}{b_a}} & 
\text{for }b_a \neq 0  \; .
\end{cases}
\end{multline}
As we will see, with the particle content of the model, the 
one-loop beta function coefficients are $b_3 = 0, b_2 = 4$ and 
$b_1 = 36/5$ (in the $\SU(5)$ normalization), 
just as in Ref.~\cite{Kribs:2007ac}. 
In particular, assuming that there are no additional fields with 
standard model charges between the EWSB scale and the Planck scale, 
one finds: 
\begin{align}
  \label{eq:ren-output}
  w_3 \left(1\tev \right) &\approx 5.4 \times  
              w_3 \left(\MP \right)  \\
  w_2 \left(1\tev \right) &=  
              w_2 \left(\MP \right)  \\
  w_1 \left(1\tev \right) &\approx  0.58 \times
              w_1 \left(\MP \right)
\end{align}

Replacing $\Phi$ by $\Sigma_a$ in Eqs.~(\ref{op2}-\ref{eq:scalar}) one can
find the gravity mediated operators involving the $\Sigma_a$
fields. The part of the Lagrangian in our model which contains these
adjoint fields  consists of the following operators:
\begin{equation}
 \begin{split}
  \label{eq:int-sigma}
 & \int\!\! d^4\!\theta \ \sum_a \Big( 1 + c_{a} 
      \frac{X^\dag X}{\MP^2} \Big) \: \text{Tr} \left[ 
            \mathbf{e}^{- g_a V_a }   \Sigma_a^{\dag} 
               \mathbf{e}^{ g_a V_a } \:\Sigma_a \right] \\
 & \qquad \qquad \qquad
 + \int\!\! d^4\!\theta \ \sum_a  k_a \: \frac{X^\dag X}{\MP^2} 
     \text{Tr} \: \Sigma_a^{2} \; .
  \end{split}
\end{equation}
Squared soft masses for the $\Sigma$ fields are generated from
operators with dimensionless coupling constants $c_a$, whereas the 
operators with couplings $k_a$ give rise to $B$-type masses.     
In the absence of renormalizable superpotential interactions of
$\Sigma$, the couplings $c_a$ do not renormalize but $k_a$ do. 
\begin{align}
  c_{a} \left(\mu \right) & \ = \ c_a \left(\MP \right) \\
  k_{a}\left(\mu \right) & \ = \ k_{s} \left(\MP 
             \right) \Bigg( \frac{Z_{\Sigma_a} \left(\MP \right)}  
                                 {Z_{\Sigma_a} \left(\mu \right)} 
                     \Bigg) 
  \label{eq:ren-b}
\end{align}

The Dirac gaugino mass operators in Eq.~\eqref{eq:int-gauge} also
contribute to the squared mass of the $\Sigma$ fields. When expanded
in components, they give tree level masses for the real parts of
$\Sigma_a$. Once the Dirac gauginos are integrated out, the imaginary
components of $\Sigma_a$ also get masses due to one loop finite
contributions. In summary, the dominant contributions to the 
scalar mass squareds can be summarized as: 
\begin{align}
  m^2_{\text{Re}[{\Sigma_a}]}  &=  2 | w_a |^2 \frac{\avg{D}^2}{\MP^2} + 
       c_a \frac{|\avg{F_X}|^2}{\MP^2}
  \label{eq:mass_re_ad} \\
  m^2_{\text{Im}[{\Sigma_a}]} &=  \frac{N_a\alpha_a}{\pi}
       \log \left( \frac{m^2_{R_a}}{M^2_{a}} \right) 
          | w_a |^2 \frac{\avg{D}^2}{\MP^2} + 
       c_a \frac{|\avg{F_X}|^2}{\MP^2}\; 
  \label{eq:mass_im_ad}
\end{align}
Here $M_{a},m_{R_a}$ are the {\it physical} masses of the gauginos
and the real part of $\Sigma_a$ respectively.

\subsection{Matter Sector}

The matter sector in our model comprises  quark and lepton 
superfields in three generations, just like the minimal
supersymmetric standard model (MSSM). 
We represent the field content by 
$\Phi^{(n)}_{i} \in \{ Q_i, U_i, D_i, L_i, E_i\}$, where $i$ is the
generation index. Excluding the Yukawa terms, the Lagrangian  in
the matter sector may be written down as
\begin{equation}
  \label{eq:int-matter}
  \int\!\! d^4\!\theta \ \sum_n \Big( \delta_{ij} + c^n_{ij} 
      \frac{X^\dag X}{\MP^2} \Big) \: \Phi_i^{(n) \dag} 
           \mathbf{e}^{ \left( q_n^a g_a V_a \right)} \:\Phi_j^{(n)} 
\end{equation}
where $q_n^a$ designate charges of the superfield $\Phi^{(n)}$ under the 
``$a$''th gauge group and $c_{ij}^n$ are dimensionless coupling
constants. The second term in Eq.~\eqref{eq:int-matter} 
belongs to the gravity mediated operators in Eq.~\eqref{eq:scalar}. 
No other operators in Eqs.~(\ref{op2}-\ref{eq:b}) are allowed because
of the charges of the matter fields. 

The Planck-suppressed operator coefficient matrices $c_{ij}^n$ are
in general anarchic in flavor space, and thus 
give rise to flavor-arbitrary squark and slepton masses.
Given the absence of Majorana gaugino masses, the scalar masses  
evolve with the renormalization group scale only because of the 
ordinary Yukawa couplings 
of the chiral superfields, 
\begin{equation}
   \label{eq:ren-scalar}
   \begin{split}
& c_{ij}\left(\mu \right) = c_{kl}\  \left(\MP \right) 
    \textbf{P}\exp\Big( - \int_{0}^{t} \! \! dt' 
       \frac{ \gamma_{ijkl}}{16 \pi^2} \:  \Big) 
\\ & \qquad \qquad \qquad \text{where} \\
& \gamma_{ijkl} \: = \: \frac{1}{2} \:
        y_{ipq}^* y_{kpq} \delta_{jl} +  \frac{1}{2} \:
            y_{lpq}^* y_{jpq}
        \delta_{ik}  +  2 y_{ipq}^* y_{jpq}
          \; ,
  \end{split} 
\end{equation}
where $t = \log \mu$,  $\textbf{P} \exp()$ refers to path ordered
exponential to account for any non-commutativity of the matrices 
$y^* y$.

In addition to the RG evolution of squark and slepton masses,
the Dirac gauginos induce a one-loop finite contribution to their masses. 
This generates one-loop finite flavor-universal contributions:  
\begin{equation}
  \label{eq:flav-diag}
  \Delta m^2_{ij}  \; = \;  \delta_{ij} \: c_i^0 \ \frac{\avg{D}^2}{\MP^2} 
\end{equation}
where
\begin{equation}
  c_i^0 \; = \; \sum_a
     \casim\left(\textbf{R}_i\right) \frac{\alpha_a}{\pi} \ w_a^2
      \ \log \left( \frac{m^2_{R_a}}{M^2_{a}} \right) \; . 
\end{equation}
Here $M_{a},m_{R_a}$ are the physical masses of the gauginos
and the real part of the scalar gaugino respectively, and 
$\casim\left(\textbf{R}_i\right)$ is the quadratic Casimir of
the representation $\textbf{R}_i$.
Therefore, the scalar soft masses at the EWSB scale are given as: 
\begin{equation}
  \label{eq:mass-sq}
  m^2_{ij} = \delta_{ij} \: c_i^0  \ \frac{\avg{D}^2}{\MP^2}
            + c_{ij}   \frac{|\avg{F_X}|^2}{\MP^2}
\end{equation}

\subsection{Higgs Sector}

In the minimal $R$-symmetric supersymmetric standard model~\cite{Kribs:2007ac},
the $\UR$-extended Higgs sector does not have the usual
Higgsino mass (namely, $\mu H_u H_d$).  Instead, it contains an additional 
pair of electroweak doublets $R_{u,d}$ that do not couple to matter and have
new mass terms $\mu_{u,d} H_{u,d} R_{u,d}$.  Different
$R$-charge assignments of the $R$ and $H$ fields ensure that they do
not mix among themselves. 

In contrast, we do not impose explicit $\UR$ in our model, and so 
mixed terms such as 
$\int\!\! d^4 \theta \ H^\dag e^V R$, 
$\int\!\! d^4 \theta \ X^\dag X H^\dag R$ 
can arise in the effective theory. These operators force
$R$ and $H$ fields to be rotated among themselves that ultimately
lead to the $R$-violating $\mu H_u H_d$ term and $R$-violating couplings 
of $R$ fields with the matter sector fields. 
Therefore, we need a symmetry to distinguish $H$ fields from $R$ fields.

Our proposal extends the gauge symmetry of the visible sector by an
additional $\UX$, under which $R$ transforms but $H$ does
not\footnote{A global or discrete symmetry can also be invoked to
  distinguish $H$ and $R$. However, global symmetries are not
  compatible with quantum gravity and typically, a discrete symmetry
  would  be broken spontaneously at the EWSB scale leading to
  formation of domain walls.}.
Fields in the Higgs sector of our model and their quantum numbers  are
summarized in the following table: 
\begin{equation}
\begin{tabular}{ c|c|c|c|c }
                   & $\SUC$    & $\SUL$    & $\UY$     & $\UX$    \\
\hline
$H_{u,d}$          & ${\bf 1}$ & ${\bf 2}$ & $\pm 1/2$ & $0$    \\
$R_{u,d}$          & ${\bf 1}$ & ${\bf 2}$ & $\mp 1/2$ & $\pm 1$  \\ 
$S_{u,d}$          & ${\bf 1}$ & ${\bf 1}$ & $0$       & $\mp 1$  \\ 
$T_{u,d}$          & ${\bf 1}$ & ${\bf 1}$ & $0$       & $\pm 2$  \\ 
\hline
\end{tabular}
\label{H-charges}
\end{equation}
We utilize the following set of marginal interactions to ensure  masses
for all matter and Higgs fields once electroweak symmetry is broken. 
\begin{equation}
  \begin{split}
  &  \int \!\! d^2 \! \theta\ \Bigg[ 
   y_u \: H_u Q U^c \ +  \ y_d\: H_d Q D^c 
               \ +\  y_\ell \: H_d L E^c  \ + \\
   & \qquad \qquad  \alpha_u \: S_u R_u H_u \ 
                   + \ \alpha_d \: S_d R_d H_d \ +  \\ 
   &  \qquad \qquad \qquad \frac{1}{2} \beta_u \: T_u S_u^2 
            \  + \ \frac{1}{2} \beta_d \: T_d S_d^2  
             \    \Bigg] \ + \ \cc \,.
   \end{split}
   \label{eq:marginal-op}
\end{equation}

The K\"ahler potential terms in our model involving Higgs fields 
can conveniently be written using the field 
$\Phi_{u,d}^{(n)} \in \{ H_{u,d}, R_{u,d}, S_{u,d}, T_{u,d}\}$:   
\begin{equation}
  \begin{split}
  \label{eq:int-higgs}
  & \int\!\! d^4\!\theta \ \sum_{n} \sum_{i \in \{ u,d\} } 
        \Big( 1 + c_{n} \: \frac{X^\dag X}{\MP^2} \Big) \: 
          \Phi_{i}^{(n) \dag} 
           \mathbf{e}^{ \left( q_{ni}^a V_a \right)} \:\Phi_{i}^{(n)} \\
   & \qquad \qquad \qquad        
+ \int\!\! d^4\!\theta \ \sum_n  k_n \:  \frac{X^\dag X}{\MP^2} 
         \Phi_u^{(n)} \Phi_d^{(n)}  
    \end{split} 
\end{equation}
The operator proportional to dimensionless coupling constants $c_n$ 
give rise to scalar soft masses of the Higgs fields, whereas the
ones proportional to $k_n$ generate the $B$-type masses. 
The renormalization group (RG) evolution of $c_n$ and $k_n$ are identical 
to the ones listed in Eq.~\eqref{eq:ren-scalar} and Eq.~\eqref{eq:ren-b}
respectively.

\subsection{Emergence of an accidental $\UR$}

All marginal operators written down in this section are exactly
invariant under the following $R$-charge assignment: 
\begin{multline} 
  \label{eq:R-charges}
    \qquad \qquad \Rch{Q, U, D, L, E} = 1  \qquad \Rch{W_a} = 1 \\  
     \Rch{\Sigma_a} = 0 \qquad  \Rch{H, S} = 0  \qquad \Rch{R, T} = 2
\end{multline}
This assignment is an extended version of the $\UR$ symmetry that the
authors in Ref.~\cite{Kribs:2007ac}  
imposed explicitly at the EWSB scale in order to reduce the 
contribution of flavor-arbitrary scalar soft masses 
to the flavor violating  
observables. They exploited the fact that the dominant (dimension
$5$) operators that are responsible for large loop contribution to
these observables also violate $\UR$ by two units.

It is striking to realize that all but two
contact terms in our gravity-mediated potential from the previous
subsection preserve this $\UR$, even though we
have not imposed it anywhere in our construction. Only the $b-$type
operators $\int\!\! d^4 \theta \ R_u R_d$ and 
$\int\!\! d^4 \theta \ T_u T_d$ violate $\UR$. However, note that
each of these terms violate $\UR$ by four units and since these are
the only terms that violate $\UR$, all effective operators that are
generated at the EWSB scale either preserve $\UR$ or violate $\UR$ by
units of $4, 8, 12, \dots$. The dimension$~5$ operators that put
severe constraints on arbitrary $c_{i j}$ would not be generated at
all in our model.

\subsection{CP violation in the Higgs Sector}

In our Higgs sector, new phases can appear in the superpotential 
couplings $\alpha_{u,d}$ and $\beta_{u,d}$ and in the $B$-type soft 
masses $B_{h,r,s,t}$. However, there are only two combinations that are 
invariant under arbitrary phase redefinitions of these fields:%
\begin{equation}
  I_1 = \alpha_u \alpha_d (B_h B_r B_s)^*  \>,\quad
  I_2 = \beta_u \beta_d (B_s^2 B_t)^*  \,.
\end{equation}
In addition to $I_1$, using $\alpha_{u,d}$ and $B_{h,r,s}$, we could 
also form combinations like $\alpha_u \alpha_d / (B_h B_r B_s)$ or $B_h 
B_r B_s / (\alpha_u \alpha_d)$ which would be invariant under phase 
redefinition.  However, these are not relevant for CP violation for 
the following reasons. First, physical observables must not involve a 
negative power of any of the $B$ parameters, because no field becomes
massless in the limit $B \to 0$ (as long as we keep nonzero soft 
mass-squareds for the scalars) so we should be able to expand any amplitude 
in positive powers of $B$.  Second, $1/\alpha_{u,d}$ are not allowed 
because if one of $\alpha_{u,d}$ happens to be zero, we can completely 
remove all phases from $\alpha_{u,d}$ and $B_{h,r,s}$, so any CP 
violation from this set of parameters should vanish as $\alpha_{u,d} \to 
0$.  Therefore, all of $\alpha_{u,d}$ and $B_{h,r,s}$ must appear with 
positive powers, hence $I_1$ is the unique invariant.  A similar 
argument can be made to single out $I_2$ for the $\beta$-$B_s$-$B_t$ 
sector.  Combined together, any CP violating physical observable must 
involve positive powers of $I_{1,2}$.

Furthermore, recall that with our R-charge assignment, $B_r$ and $B_t$ 
each break R-charge by 4 while all other parameters are R-neutral, so 
$I_1$ and $I_2$ each carry R-charge 4.  Then, since all CP-violating 
observables in the standard model are $R$-neutral, they must appear 
in the combination 
$I_1 I_2^*$. This is a dimension-12 object which involves all four of 
$\alpha_{u,d}$ and $\beta_{u,d}$, so CP is broken not only really softly 
but also at a very high loop order, which therefore should be 
inconsequential.

\section{Phenomenology}
\label{sec:pheno}

A comprehensive study of the parameter space in our model is beyond
the scope of this paper. We rather find one point that is allowed
under the flavor and CP violation constraints and satisfies 
other experimental bounds. 

There are several novel phenomenological features of this point. At the
weak scale, the point contains large flavor violation and Dirac
gauginos that have  been studied before in
the context of the minimal $R$-symmetric supersymmetric standard
model (see for example,~\cite{Kribs:2009zy,Herquet:2010ka} and
\cite{Kribs:2008hq}).  In addition to these, our benchmark point also
contains a light $Z'$ with notrivial couplings to the Higgs sector.

\subsection{Benchmark Point}
\label{sec:benchmark}

First, let us choose gaugino masses.  To reduce the number of parameters,
let us make a simplifying assumption that the
$\SU(3) \otimes \SU(2) \otimes \UY$
gauginos have an equal mass at the Planck scale, i.e.:
\begin{equation}
  \label{eq:input-gaugino}
  w_{3,2,1}\bigr|_{\MP} = 1 
  \quad\text{and}\quad 
  \left. \frac{\avg{D}}{\MP}\right|_{\MP} = 1\tev  \,,
\end{equation}
where $w_{1,2,3}$ are defined in Eq.~(\ref{eq:dirac}) with 
the ``$\SU(5)$ normalization'' for $\UY$, although it is not 
our intention here to imply or require grand unification.
Then, after evolving the gaugino masses with renormalization
group running Eq.~(\ref{eq:ren-dirac}) down to $\mu = 1\tev$, we find
\begin{equation}
  \label{eq:output-gaugino}
  M_3 = 5.4\tev \,,\>  M_2 = 1.0\tev  \,,\> M_1 = 580\gev \,.
\end{equation}
Instead of treating $w_X$ at $\MP$ as input parameter and then
renormalizing it down to the EWSB scale, we rather choose the mass
parameter $M_X$ at $1\tev$ to be 
\begin{equation}
  \label{eq:X-gaugino}
  M_X = 500\gev \; .
\end{equation}

\subsection{Flavor/CP Violation}
\label{sec:flav_vio}

Let us analyze the most stringent constraints, the flavor and 
$CP$ violation in $K^0$-$\overline{K}{}^0$ mixing.  When generated at the 
Planck scale, the coefficient matrix $c_{ij}$ of the K\"ahler term 
$X^\dag X Q^\dag_i Q_j$ (and similarly for $U^c$ and $D^c$) 
should be ``anarchic''; to be more precise, we take each element 
of $c_{ij}$ to have a magnitude and a phase in the range from $0.1$ to $1$.

As in Eq.~(\ref{eq:mass-sq}), the low-energy squark mass matrix 
also receives contributions from integrating out the heavy Dirac gauginos 
(dominantly the gluino).  For the benchmark gluino mass, we obtain
\begin{eqnarray}
  \label{eq:mass-sq-squark}
  m^2_{ij} &=& \delta_{ij} \: (1.3\tev)^2  \ + \ 
                  c_{ij}  \frac{|\avg{F_X}|^2}{\MP^2} \nonumber\\ 
           &=& (1.3\tev)^2 \left( \delta_{ij} \ + \ 
           c_{ij} \frac{|\avg{F_X}|^2}{\avg{D}^2} \right) .               
\end{eqnarray}
(Here, the running of $X^\dag X Q^\dag Q$ operator was neglected since it 
depends only on small Yukawa couplings at one-loop when gauginos are Dirac.
To the extent that there is RG evolution, its effect is to further lower 
the squark mass with respect to the gluino mass, only making our 
discussion stronger.) Thus, the flavor violating parameter
$\delta_{ij} \equiv \Delta m^2_{ij}/m^2_{\rm avg}$ is given by 
\begin{equation}
  \label{eq:delta}
  \delta_{ij} \approx \frac{\avg{F_X}^2}{\avg{D}^2} \frac{c_{ij}}{c^0} 
                      \quad \text{for } i\neq j.
\end{equation}

The strongest constraint from $\Delta F = 2$ observables follow from 
$K^0$-$\overline{K}{}^0$ mixing 
($\delta_{L} = \delta_{R} \leq 0.08$) for gluino and squark
masses to be order of $5\tev$ and $1\tev$
respectively~\cite{Blechman:2008gu}).   
Assuming that $\avg{D} \sim 1.5~|\avg{F_X}| $,  we find that  
\begin{equation}
  \label{eq:sq-kkbar}
  c^{q}_{ij} \ \leq \ 0.3 \times 
  \left( \begin{array}{ccc}
  \sim 1 & \sim 1 & \sim 1 \\
  \sim 1 & \sim 1 & \sim 1 \\
  \sim 1 & \sim 1 & \sim 1 
  \end{array} \right) \; .
\end{equation}
An even stronger bound arises due to 
CP violation in $K^0$-$\overline{K}{}^0$ mixing. 
Roughly, the CP violating quantity
$\text{Im}(\delta_L^* \delta_R) \lsim 10^{-3}$,
which can be accomplished by taking $c^q_{ij} \leq 0.1$
(multiplying an anarchic matrix) as well as the
relative phase to be smaller than $0.1$.  

Therefore, we conclude that it is possible to satisfy the 
stringent bound from the $CP$ violation in $K^0$-$\wba{K^0}$ 
even with quite ``flavorful'' sfermion mass-squareds at the Planck scale, 
given the ratio $m_\text{squark} / m_\text{gluino} \sim \cO(0.1)$,
which arises naturally from RG evolution in our model.
As discussed in Refs.~\cite{Kribs:2007ac,Blechman:2008gu},
all other squark flavor bounds are satisfied once this ratio is assumed.  

Lepton flavor violation also provides constraints on
the slepton mass matrices~\cite{Kribs:2007ac,Fok:2010vk}.
Maximal mixing is strongly constrained, particularly in the 
right-handed slepton sector, due to constraints from both 
$\mu \ra e$ conversion as well as $\mu \ra e\gamma$~\cite{Fok:2010vk}.  
For a Dirac bino of mass $600\gev$, a relatively mild restriction on the 
mixing angle $\sin 2\theta_{l} \lsim 0.3 \ra 0.5$ for nondegenerate 
slepton mass eigenstates is necessary.  
For right-handed sleptons, since the dominant contribution to 
their masses is from the flavor-arbitrary K\"ahler potential
operators, this implies some modest degeneracy needed, 
roughly $c_{12}^{l}/c_{ii}^{l} \lsim 0.15 \ra 0.3$.
This is well within our stated goal of coupling constants
remaining within the range $0.1 \ra 1$.

\subsection{Electroweak and $\UX$ Breaking}
\label{sec:EW_and_U1X_breaking}

We seek a minimum of the scalar potential where 
$\avg{H_{u,d}} \neq 0$
and $\avg{S_{u,d}} \neq 0$ and $\avg{R_{u,d}} = \avg{T_{u,d}} = 0$, so
that the electroweak symmetry is broken down to electromagnetism
($\SUL \times \UY \rightarrow \U(1)_\text{EM}$) and $\UX$ is broken
completely, while preserving the ``accidental'' $\UR$ of the visible
sector.  The potential in 
our model depends on soft masses of all the
multiplets in the Higgs sector (i.e.~$H_{u,d}, R_{u,d},
S_{u,d}$ and $T_{u,d}$), soft masses of the adjoints $\Sigma_a$
and the Yukawa couplings $\alpha_{u,d}$ and $\beta_{u,d}$.

Let us first consider the effect of the scalar adjoints.
As shown in Eq.~\eqref{eq:mass_re_ad}, the real parts of the scalar  
adjoints receive large masses from Dirac gaugino mass operators. 
For our benchmark scenario, their physical masses are at or above
$1\tev$, while we expect the mass parameters in the Higgs sector are 
typically of the order of the weak scale.
Hence, we integrate out the scalar adjoints to analyze 
the scalar potential in the Higgs sector. 

In this limit, as is well known, the quartic terms in the Higgs-$\UX$
sector vanish in the absence of soft masses  
$m_{\tilde{\Sigma}_a}^2$ and $b_{\tilde{\Sigma}_a^2}$. 
Once soft masses for these scalars are generated, 
the quartic is partially restored. In
particular the $D$-term due to each gauge group shifts according to: 
\begin{equation}
  \label{eq:gauge-quartic}
  \frac{g_a^2}{8} \ \rightarrow \ \eta_a   \frac{g_a^2}{8} \  =  \ 
  \left( \frac{ m_{\Sigma_a}^2 -
      b_{\Sigma_a} }{ 4 M_a^2 + m_{\Sigma_a}^2 +
      b_{\Sigma_a}} \right)  \  \frac{g_a^2}{8}
\end{equation}
With our gaugino masses and using 
$m_{\Sigma_a}^2 \sim - b_{\Sigma_a} \sim (670\gev)^2$ 
(it corresponds to $c_{\Sigma_a} \sim - k_{\Sigma_a} \sim 1 $), we
find that the MSSM $D$-term effectively scales approximately as
$(g_1^2+g_2^2)/8 \rightarrow (1/4)  \ (g_1^2+g_2^2)/8$. 
This is not meant to be the best or even indicative of the
quartic suppression -- it simply illustrates the actual 
suppression for the particular benchmark point that we have
chosen to study.  Finally, for the $\UX$ sector we use $\eta_X = 1/3$.  

Four nontrivial conditions to attain the right vacuum structure follow 
from the minimization conditions, 
$\partial V/ \partial H_{u,d}^0 = \partial V/ \partial S_{u,d} = 0$, 
which we use to eliminate four input parameters, 
$m_{H_{u,d}}^2$, $m_{S_{u,d}}^2$ in terms of other soft and 
supersymmetry breaking parameters. In particular, we choose the following
point (all parameters in the EWSB scale):
\begin{equation}
  \label{eq:param-in}
 \begin{split}
&  \avg{S_u}^2 +  \avg{S_d}^2 \: = \: v_s^2 \: = \: 
       m_X^2/(2 g_X^2) \: = \: (600\gev)^2  \, ,  \\
&  \frac{\avg{S_u}}{\avg{S_d}} \: = \: 
           \tan\beta_X = 0.2 \>, \qquad 
  \frac{\avg{H_u^0}}{\avg{H_d^0}} = \tan\beta = 10 \,,  \\
&      \qquad \qquad \qquad  m_X = 150\gev  \\
&  \quad B_{h,r,t} =  (300\gev)^2 \>,\quad  B_s =  (500\gev)^2 \,,
        \\
& \quad  \alpha_u = 1 \>,\quad \alpha_d = 0.3 \>,\quad 
  \beta_u = 0.9\>,  \quad \beta_d = 0.7 \, , \\
&  \qquad m^2_{R_{u,d}} \ =\  m^2_{T_{u,d}} \ = \ (700\gev)^2  \, . 
 \end{split}
\end{equation}
(We have checked that these couplings do not hit Landau poles below
$\MP$.) The gauge coupling constant $g_\text{X}$ can be inferred from
the gauge boson mass $m_\text{X}$ and total $\UX$ breaking vev
$v_s$. With our parameters in Eq.~\eqref{eq:param-in} we find 
$g_X = 0.18$. These values as well as the gaugino masses 
Eqs.~(\ref{eq:output-gaugino},\ref{eq:X-gaugino})
for all gauge groups define our benchmark point.

Now, at this benchmark point, the chargino and neutralino mass
matrices are completely determined at tree level from which we find
that the lightest two neutralinos have masses of $54,176\gev$, and the 
lightest chargino has a mass of $117\gev$.  These are safely above 
the LEP limits.  

The CP-even neutral Higgs mass matrix is more complicated, where the
scalar components of $H_u^0, H_d^0, S_u, S_d$ mix among themselves. 
Naively, it might appear challenging to get Higgs masses above the 
LEP limit, given the $\eta$-suppressions of the quartic couplings, 
Eq.~(\ref{eq:gauge-quartic}).
For example, the lightest mass eigenstate in the CP-even sector 
would attain only $\simeq 20\gev$ mass at tree level at our benchmark
point. Therefore, as in the MSSM, the one-loop quantum corrections 
to the quartic coupling from top/stop loops are very important.
At our benchmark point, there are additional important one-loop contributions 
from the superpotential term proportional to the $\alpha_u$ coupling.
The radiative corrections to the Higgs quartic can be roughly 
estimated as 
\begin{equation}
  \label{eq:loop-quart}
  \frac{3 y_t^4}{32 \pi^2} \log 
      \!\left( \frac{m_{\tilde t_1}m_{\tilde t_2}} {m_t^2} \right) + 
   \frac{\alpha_u^4}{32 \pi^2} \log\! 
      \left( \frac{m_{R_u}m_{S_u}}{m_{\tilde R} m_{\tilde S}} \right)  \; .
\end{equation}
The first term in Eq.~\eqref{eq:loop-quart} is due to the familiar
top-stop mass splitting in the MSSM and the second contribution is due
to the splittings among the neutralinos and the neutral scalars in
$R_u$ and $S_u$. Strictly speaking, $\alpha_u$ in
Eq.~\eqref{eq:loop-quart} should have been replaced by
the Higgs couplings of various neutralino/neutral scalar mass
eigenstates. We can, howver, give a rough estimate of Higgs mass. 
For our benchmark point, stop masses are
$\sim 1.3\tev$ and the top-stop splitting generates $\sim 96\gev$ for
Higgs mass. Approximating $m_{\tilde{R},\tilde{S}} \sim 100\gev$
we see that $m_{R_u,S_u} \sim 1\tev$ is sufficient to obtain 
a lightest Higgs mass above the LEP bound. 

We should emphasize that the difficulty to get the Higgs mass
above the LEP bound is for our benchmark point, and does not
in general apply to our model.  There are at least two interesting ways 
to raise the Higgs mass beyond what we considered above:  
One is to raise the the scalar masses $m_{\tilde{R},\tilde{T}}$, 
to increase the one-loop contribution.  The second is to increase 
the masses of the adjoint scalars $m_{\tilde{\Sigma}_{1,2}}$, 
to increase the tree-level quartic coupling.  
Nevertheless, the lightest Higss boson does generically tend
be rather light, close to the LEP bound, given the scales
in the model.

\subsection{$\UR$ violation from anomaly mediation}

Since the gravitino mass necessarily breaks $\UR$, there are small
but irreducible contributions to Majorana gaugino masses, 
slightly violating the accidental $\UR$ of the visible sector.  
Fortunately, they do not upset our flavor protection mechanism.
This issue was already studied in Ref.~\cite{Kribs:2007ac}, which we 
briefly recast here.

The anomaly-mediated contributions to the Majorana-type 
gaugino masses are~\cite{Giudice:1998xp,Randall:1998uk}, 
in our model, 
\begin{eqnarray}
\tilde{M}_a &=& b_a \frac{\alpha_a}{4\pi} m_{3/2} \; \simeq \; \left\{
\begin{array}{ll}
0.01 m_{3/2}    & \quad a = 1 \\
0.01 m_{3/2}    & \quad a = 2 \\
0               & \quad a = 3 \; ,
\end{array} \right. 
\end{eqnarray}
where we have evaluated the $\beta$-functions for our model
content with couplings at the weak scale.
Interestingly, the anomaly-mediation-induced Majorana mass 
for the gluino is zero at one-loop because the 
$\SU(3)$ $\beta$-function vanishes for our field content.
Indeed, any visible sector model minimally extended to become 
$R$-symmetric has this property.  In practice, two-loop contributions 
to the Majorana gluino mass are present, as well as one-loop
threshold contributions resulting from differences between
the masses of the Dirac gluino versus the color octet scalars.
These are expected to lead to a contribution smaller than 
about $10^{-3} m_{3/2}$.  

The size of the gravitino mass $m_{3/2}$ relative to the visible 
sector is, in general, sensitive to the hidden sector%
~\cite{Cohen:2006qc,Roy:2007nz,Murayama:2007ge,Perez:2008ng}.
In our case, however, this effect is small.  Noting
that $m_{3/2}$ is related to the cosmological constant~\cite{Deser:1977uq} 
which is, in turn, approximately equal to the expectation value of the total
potential, we find that for our benchmark point the gravitino mass is
given as:
\begin{equation}
  \label{eq:m_32}
  m_{3/2} \ \approx \ 730\gev \; .
\end{equation}
This implies the Majorana masses for all of the gauginos 
are below about $10\gev$\@.  The small splittings in the gaugino sector
do not induce excessive $R$-violation back into flavor violating
processes.  They are, however, quite relevant to 
determining the identity of the LSP, its relic abundance,
and the associated collider phenomenology.

\subsection{$\UX$ phenomenology}

The $\UX$ gauge boson $Z'$ can be potentially very interesting given 
its rather light mass (in Eq.~(\ref{eq:param-in}) at the benchmark point).
How can we produce $Z'$ at colliders?
At tree-level, $\UX$ only couples to $S_{u,d}$ and $T_{u,d}$,
and the $S_{u,d}$ scalars can mix with $H_{u,d}$ with large mixing angles.
Therefore, one of the more important production channels is 
$gg \to h^* \to Z'+Z'$.

At one-loop level, kinetic mixing of $\UX$ and $\UY$ is inevitably 
induced from the loops of $R_{u,d}$, which carry both charges.  
Therefore, through this kinetic mixing, the $Z'$ emitted off the Higgs 
will in turn decay to a pair of standard model fermions.  
We leave a detailed analysis 
of this interesting signal and other potentially useful channels 
for probing $\UX$ to future work.  Here, we check that such a light 
$Z'$ is allowed by existing constraints.

First, note that the $\UX$-$\UY$ kinetic mixing is much smaller 
compared to the analogous mixing for the well-studied extension
involving $\U(1)_{B-L}$.  This is because 
the only fields running in the loop in our case 
are $R_{u,d}$, compared to the entire matter content of the standard model
in the $B-L$ case.
More explicitly, the mixing coefficient can be estimated as%
\begin{equation}
\begin{split}
  \epsilon \ &= \  \sum_i q^i_Y q^i_X \  \frac{g_Y g_X}{8\pi^2}\  
            \log \frac{\Lambda}{m_i} \\
             &\approx \ 1.6 \times 10^{-3} \ 
             \left( \frac{g_X}{0.18} \right) \     
            \log \frac{\Lambda}{1\tev}
\end{split}
\end{equation}
where $q^i_Y$ and $q^i_X$ are charges of field $i$ under $\UY$
and $\UX$ respectively and $m_i$ is the mass. The scale $\Lambda$ 
is some high mass scale above which there is no mixing 
({\it e.g.} embedding $\UX$ into a non-Abelian group). 

The kinetic mixing implies the $Z'$ couples to all standard model 
matter particles and also contributes to precision electroweak observables.  
The strength of the $Z'$ coupling to standard model fields is given by
$q^i_Y \epsilon  g_Y$, where $q^i_Y$ are the hypercharges of
corresponding standard model fields.  The bound on the 
$Z'$ mass can be directly obtained from Ref.~\cite{Cacciapaglia:2006pk}: 
$m_X \: \gtrsim \ \epsilon \: g_Y \times 6.7\tev$\@. 

Considering only the $Z'$ couplings proportional to $\epsilon$, 
an interesting generalized bound was recently analyzed in 
Ref.~\cite{Hook:2010tw}.  For $\epsilon \lesssim 3\times 10^{-2}$, 
$m_X$ is allowed to be in the entire range between
$10-1000\gev$\@.  For our benchmark point, we find that 
$\epsilon$ would exceed $3\times 10^{-2}$ only when $\Lambda$ exceeds
$10^{11}\gev$.  The precise bounds in our model requires 
further study, however.  This is because the $Z'$-stralung process
is undoubtedly constrained for small $Z'$ masses.  We will return to
this very interesting phenomenological issue in future work.

\section{Unification, Unifons, and Singlets}
\label{unifon-sec}

As pointed out in Ref.~\cite{Kribs:2007ac}, gauge coupling unification
of the standard model gauge couplings is not automatic.
The $\SU(3)$ color and $\SU(2)$ weak interactions do continue to unify 
to a perturbative value near $10^{16}\gev$  as in the MSSM, since 
the shift in the $\beta$-functions from the enlarged field content 
($\Sigma_3,\Sigma_2$,$R_u$,$R_d$) is equivalent to three sets 
of fundamentals and antifundamentals for $\SU(3)$ and $\SU(2)$.  
The $\UY$ coupling $\beta$-function, however, is only shifted
by one unit.  Hence, two pairs of vector-like unifons with
hypercharge $q_Y = \pm 1$ transforming under $\UY$ are 
sufficient to have the standard model couplings unify near
$10^{16}\gev$.  

However, this is not the end of the story.  There are also 
dangerous operators in supersymmetric models involving 
singlets~\cite{Bagger:1995ay}.
Dirac gaugino masses for $\UY$ and $\UX$ require
gauge singlets in the visible sector.  A completely general 
K\"ahler potential would allow operators such as
\begin{eqnarray}
\int d^4 \theta \, \frac{X^\dagger X}{\MP} \Sigma_{Y,X}
\end{eqnarray}
that lead to large tadpoles for the singlet fields $\Sigma_{Y,X}$.

There are two possibilities for dealing with this.  One is to
assume that the $\U(1)$s are embedded into one or more non-Abelian 
(unified) groups.  So long as this unification is not too close 
to the Planck scale, higher dimensional operators involving
breaking fields of the unified sector would keep the coefficients
of the tadpole operators small enough.

Another possibility is to not have singlets at all in the
visible sector.  This would mean both $\UY$ and 
$\UX$ would acquire purely Majorana-type masses.  
While this may be a viable solution for $\UX$,
a pure Majorana bino would reintroduce large lepton 
flavor violation.  The most dangerous lepton flavor violation
would arise in operators involving the right-handed sleptons,
since their masses are (in general) entirely anarchic.
The naive bounds on $\delta_{RR}$ suggest tuning at level
of $10^{-2}$ would be needed.

\section{Discussion}
\label{discussion-sec}

We have presented a complete model of gravity-mediated supersymmetry
breaking with a visible sector that is safe from flavor constraints.
We exploited the observation that flavor violation beyond the
standard model can be greatly suppressed 
\emph{even without flavor universality of the squark and slepton sectors} 
if the visible sector particle content respects an approximate global $\UR$
symmetry~\cite{Kribs:2007ac}.  
The key difference between previous approaches to $R$-symmetry
and the one pursued here is that the origin of the approximate 
$\UR$ symmetry is \emph{accidental}.  Since we employ gravity-mediation,
all operators must be RG evolved to the weak scale, and this 
causes a large upward renormalization of the gluino mass.
In this way, the ``little hierarchy'' in which the Dirac
gluino mass was needed to be higher than the squark masses
is automatic in our framework.  

We emphasize that all flavor problems are solved with our
framework, including $\Delta m_K$, $\epsilon_K$, and 
lepton flavor violation constraints including
$\mu \ra e$ conversion and $\mu \ra e\gamma$.
The solution can be mostly attributed to $R$-symmetry and 
the large renormalization of the gluino mass.  
This in itself not quite enough~\cite{Blechman:2008gu},
and so we must also allow ourselves to tune $c^{\tilde{q}}_{ij}$ 
coefficients to of order $0.1$ in magnitude and phase
to fully comply with the constraint from $\epsilon_K$.
Somewhat more relaxed restrictions on the $c^{l}_{ij}$
maintain safely from lepton flavor violation constraints.

Our philosophy exploited holomorphy of the superpotential while we
strictly eliminated dangerous nonchiral K\"ahler terms through 
an additional gauged $\UX$ symmetry in the Higgs sector.
The phenomenology of this $\UX$ is quite interesting:  
the $Z'$ of this broken $\UX$ couples with order one strength 
to Higgs fields and order $10^{-4} \ra 10^{-2}$ strength to 
particles carrying hypercharge.  More detailed phenomenology
is left to future work, but clearly the prospects for detection
at LHC and Tevatron are bright!

\section*{Acknowledgments}

We thank Kaustubh Agashe, Cedric Delaunay, Paddy Fox and Adam Martin
for discussions. GDK thanks the Aspen Center for Physics and the
Fermilab Theoretical Physics Group for hospitality 
where part of this work was done.  
GDK and TSR were supported in part by the US Department of Energy 
under contract number DE-FG02-96ER40969.
TO was supported in part by a First Year Assistant Professor Award
and the Department of Physics at Florida State University.



\begin{thebibliography}
\expandafter\ifx\csname natexlab\endcsname\relax\def\natexlab#1{#1}\fi
\expandafter\ifx\csname bibnamefont\endcsname\relax
  \def\bibnamefont#1{#1}\fi
\expandafter\ifx\csname bibfnamefont\endcsname\relax
  \def\bibfnamefont#1{#1}\fi
\expandafter\ifx\csname citenamefont\endcsname\relax
  \def\citenamefont#1{#1}\fi
\expandafter\ifx\csname url\endcsname\relax
  \def\url#1{\texttt{#1}}\fi
\expandafter\ifx\csname urlprefix\endcsname\relax\def\urlprefix{URL }\fi
\providecommand{\bibinfo}[2]{#2}
\providecommand{\eprint}[2][]{\url{#2}}

\bibitem[{\citenamefont{Chamseddine et~al.}(1982)\citenamefont{Chamseddine,
  Arnowitt, and Nath}}]{Chamseddine:1982jx}
\bibinfo{author}{\bibfnamefont{A.~H.} \bibnamefont{Chamseddine}},
  \bibinfo{author}{\bibfnamefont{R.~L.} \bibnamefont{Arnowitt}},
  \bibnamefont{and} \bibinfo{author}{\bibfnamefont{P.}~\bibnamefont{Nath}},
  \bibinfo{journal}{Phys. Rev. Lett.} \textbf{\bibinfo{volume}{49}},
  \bibinfo{pages}{970} (\bibinfo{year}{1982}).

\bibitem[{\citenamefont{Barbieri et~al.}(1982)\citenamefont{Barbieri, Ferrara,
  and Savoy}}]{Barbieri:1982eh}
\bibinfo{author}{\bibfnamefont{R.}~\bibnamefont{Barbieri}},
  \bibinfo{author}{\bibfnamefont{S.}~\bibnamefont{Ferrara}}, \bibnamefont{and}
  \bibinfo{author}{\bibfnamefont{C.~A.} \bibnamefont{Savoy}},
  \bibinfo{journal}{Phys. Lett.} \textbf{\bibinfo{volume}{B119}},
  \bibinfo{pages}{343} (\bibinfo{year}{1982}).

\bibitem[{\citenamefont{Ibanez}(1982)}]{Ibanez:1982ee}
\bibinfo{author}{\bibfnamefont{L.~E.} \bibnamefont{Ibanez}},
  \bibinfo{journal}{Phys. Lett.} \textbf{\bibinfo{volume}{B118}},
  \bibinfo{pages}{73} (\bibinfo{year}{1982}).

\bibitem[{\citenamefont{Hall et~al.}(1983)\citenamefont{Hall, Lykken, and
  Weinberg}}]{Hall:1983iz}
\bibinfo{author}{\bibfnamefont{L.~J.} \bibnamefont{Hall}},
  \bibinfo{author}{\bibfnamefont{J.~D.} \bibnamefont{Lykken}},
  \bibnamefont{and} \bibinfo{author}{\bibfnamefont{S.}~\bibnamefont{Weinberg}},
  \bibinfo{journal}{Phys. Rev.} \textbf{\bibinfo{volume}{D27}},
  \bibinfo{pages}{2359} (\bibinfo{year}{1983}).

\bibitem[{\citenamefont{Ohta}(1983)}]{Ohta:1982wn}
\bibinfo{author}{\bibfnamefont{N.}~\bibnamefont{Ohta}}, \bibinfo{journal}{Prog.
  Theor. Phys.} \textbf{\bibinfo{volume}{70}}, \bibinfo{pages}{542}
  (\bibinfo{year}{1983}).

\bibitem[{\citenamefont{Ellis et~al.}(1983)\citenamefont{Ellis, Nanopoulos, and
  Tamvakis}}]{Ellis:1982wr}
\bibinfo{author}{\bibfnamefont{J.~R.} \bibnamefont{Ellis}},
  \bibinfo{author}{\bibfnamefont{D.~V.} \bibnamefont{Nanopoulos}},
  \bibnamefont{and} \bibinfo{author}{\bibfnamefont{K.}~\bibnamefont{Tamvakis}},
  \bibinfo{journal}{Phys. Lett.} \textbf{\bibinfo{volume}{B121}},
  \bibinfo{pages}{123} (\bibinfo{year}{1983}).

\bibitem[{\citenamefont{Alvarez-Gaume et~al.}(1983)\citenamefont{Alvarez-Gaume,
  Polchinski, and Wise}}]{AlvarezGaume:1983gj}
\bibinfo{author}{\bibfnamefont{L.}~\bibnamefont{Alvarez-Gaume}},
  \bibinfo{author}{\bibfnamefont{J.}~\bibnamefont{Polchinski}},
  \bibnamefont{and} \bibinfo{author}{\bibfnamefont{M.~B.} \bibnamefont{Wise}},
  \bibinfo{journal}{Nucl. Phys.} \textbf{\bibinfo{volume}{B221}},
  \bibinfo{pages}{495} (\bibinfo{year}{1983}).

\bibitem[{\citenamefont{Nilles}(1984)}]{Nilles:1983ge}
\bibinfo{author}{\bibfnamefont{H.~P.} \bibnamefont{Nilles}},
  \bibinfo{journal}{Phys. Rept.} \textbf{\bibinfo{volume}{110}},
  \bibinfo{pages}{1} (\bibinfo{year}{1984}).

\bibitem[{\citenamefont{Nath et~al.}(1984)\citenamefont{Nath, Arnowitt, and
  Chamseddine}}]{Nath:1983fp}
\bibinfo{author}{\bibfnamefont{P.}~\bibnamefont{Nath}},
  \bibinfo{author}{\bibfnamefont{R.~L.} \bibnamefont{Arnowitt}},
  \bibnamefont{and} \bibinfo{author}{\bibfnamefont{A.~H.}
  \bibnamefont{Chamseddine}} (\bibinfo{year}{1984}), \bibinfo{note}{lectures
  given at Summer Workshop on Particle Physics, Trieste, Italy, Jun 20 - Jul
  29, 1983}.

\bibitem[{\citenamefont{Giudice and Masiero}(1988)}]{Giudice:1988yz}
\bibinfo{author}{\bibfnamefont{G.~F.} \bibnamefont{Giudice}} \bibnamefont{and}
  \bibinfo{author}{\bibfnamefont{A.}~\bibnamefont{Masiero}},
  \bibinfo{journal}{Phys. Lett.} \textbf{\bibinfo{volume}{B206}},
  \bibinfo{pages}{480} (\bibinfo{year}{1988}).

\bibitem[{\citenamefont{Dimopoulos and Georgi}(1981)}]{Dimopoulos:1981zb}
\bibinfo{author}{\bibfnamefont{S.}~\bibnamefont{Dimopoulos}} \bibnamefont{and}
  \bibinfo{author}{\bibfnamefont{H.}~\bibnamefont{Georgi}},
  \bibinfo{journal}{Nucl. Phys.} \textbf{\bibinfo{volume}{B193}},
  \bibinfo{pages}{150} (\bibinfo{year}{1981}).

\bibitem[{\citenamefont{Dimopoulos
  et~al.}(1981{\natexlab{a}})\citenamefont{Dimopoulos, Raby, and
  Wilczek}}]{Dimopoulos:1981yj}
\bibinfo{author}{\bibfnamefont{S.}~\bibnamefont{Dimopoulos}},
  \bibinfo{author}{\bibfnamefont{S.}~\bibnamefont{Raby}}, \bibnamefont{and}
  \bibinfo{author}{\bibfnamefont{F.}~\bibnamefont{Wilczek}},
  \bibinfo{journal}{Phys. Rev.} \textbf{\bibinfo{volume}{D24}},
  \bibinfo{pages}{1681} (\bibinfo{year}{1981}{\natexlab{a}}).

\bibitem[{\citenamefont{Gabbiani and Masiero}(1989)}]{Gabbiani:1988rb}
\bibinfo{author}{\bibfnamefont{F.}~\bibnamefont{Gabbiani}} \bibnamefont{and}
  \bibinfo{author}{\bibfnamefont{A.}~\bibnamefont{Masiero}},
  \bibinfo{journal}{Nucl. Phys.} \textbf{\bibinfo{volume}{B322}},
  \bibinfo{pages}{235} (\bibinfo{year}{1989}).

\bibitem[{\citenamefont{Gabbiani et~al.}(1996)\citenamefont{Gabbiani,
  Gabrielli, Masiero, and Silvestrini}}]{Gabbiani:1996hi}
\bibinfo{author}{\bibfnamefont{F.}~\bibnamefont{Gabbiani}},
  \bibinfo{author}{\bibfnamefont{E.}~\bibnamefont{Gabrielli}},
  \bibinfo{author}{\bibfnamefont{A.}~\bibnamefont{Masiero}}, \bibnamefont{and}
  \bibinfo{author}{\bibfnamefont{L.}~\bibnamefont{Silvestrini}},
  \bibinfo{journal}{Nucl. Phys.} \textbf{\bibinfo{volume}{B477}},
  \bibinfo{pages}{321} (\bibinfo{year}{1996}), \eprint{hep-ph/9604387}.

\bibitem[{\citenamefont{Bagger et~al.}(1997)\citenamefont{Bagger, Matchev, and
  Zhang}}]{Bagger:1997gg}
\bibinfo{author}{\bibfnamefont{J.~A.} \bibnamefont{Bagger}},
  \bibinfo{author}{\bibfnamefont{K.~T.} \bibnamefont{Matchev}},
  \bibnamefont{and} \bibinfo{author}{\bibfnamefont{R.-J.} \bibnamefont{Zhang}},
  \bibinfo{journal}{Phys. Lett.} \textbf{\bibinfo{volume}{B412}},
  \bibinfo{pages}{77} (\bibinfo{year}{1997}), \eprint{hep-ph/9707225}.

\bibitem[{\citenamefont{Ciuchini et~al.}(1998)}]{Ciuchini:1998ix}
\bibinfo{author}{\bibfnamefont{M.}~\bibnamefont{Ciuchini}}
  \bibnamefont{et~al.}, \bibinfo{journal}{JHEP} \textbf{\bibinfo{volume}{10}},
  \bibinfo{pages}{008} (\bibinfo{year}{1998}), \eprint{hep-ph/9808328}.

\bibitem[{\citenamefont{Barbieri et~al.}(1996)\citenamefont{Barbieri, Dvali,
  and Hall}}]{Barbieri:1995uv}
\bibinfo{author}{\bibfnamefont{R.}~\bibnamefont{Barbieri}},
  \bibinfo{author}{\bibfnamefont{G.~R.} \bibnamefont{Dvali}}, \bibnamefont{and}
  \bibinfo{author}{\bibfnamefont{L.~J.} \bibnamefont{Hall}},
  \bibinfo{journal}{Phys. Lett.} \textbf{\bibinfo{volume}{B377}},
  \bibinfo{pages}{76} (\bibinfo{year}{1996}), \eprint{hep-ph/9512388}.

\bibitem[{\citenamefont{Barbieri et~al.}(1997)\citenamefont{Barbieri, Hall,
  Raby, and Romanino}}]{Barbieri:1996ww}
\bibinfo{author}{\bibfnamefont{R.}~\bibnamefont{Barbieri}},
  \bibinfo{author}{\bibfnamefont{L.~J.} \bibnamefont{Hall}},
  \bibinfo{author}{\bibfnamefont{S.}~\bibnamefont{Raby}}, \bibnamefont{and}
  \bibinfo{author}{\bibfnamefont{A.}~\bibnamefont{Romanino}},
  \bibinfo{journal}{Nucl. Phys.} \textbf{\bibinfo{volume}{B493}},
  \bibinfo{pages}{3} (\bibinfo{year}{1997}), \eprint{hep-ph/9610449}.

\bibitem[{\citenamefont{Kallosh et~al.}(1995)\citenamefont{Kallosh, Linde,
  Linde, and Susskind}}]{Kallosh:1995hi}
\bibinfo{author}{\bibfnamefont{R.}~\bibnamefont{Kallosh}},
  \bibinfo{author}{\bibfnamefont{A.~D.} \bibnamefont{Linde}},
  \bibinfo{author}{\bibfnamefont{D.~A.} \bibnamefont{Linde}}, \bibnamefont{and}
  \bibinfo{author}{\bibfnamefont{L.}~\bibnamefont{Susskind}},
  \bibinfo{journal}{Phys. Rev.} \textbf{\bibinfo{volume}{D52}},
  \bibinfo{pages}{912} (\bibinfo{year}{1995}), \eprint{hep-th/9502069}.

\bibitem[{\citenamefont{Hall and Randall}(1990)}]{Hall:1990ac}
\bibinfo{author}{\bibfnamefont{L.~J.} \bibnamefont{Hall}} \bibnamefont{and}
  \bibinfo{author}{\bibfnamefont{L.}~\bibnamefont{Randall}},
  \bibinfo{journal}{Phys. Rev. Lett.} \textbf{\bibinfo{volume}{65}},
  \bibinfo{pages}{2939} (\bibinfo{year}{1990}).

\bibitem[{\citenamefont{Nir and Seiberg}(1993)}]{Nir:1993mx}
\bibinfo{author}{\bibfnamefont{Y.}~\bibnamefont{Nir}} \bibnamefont{and}
  \bibinfo{author}{\bibfnamefont{N.}~\bibnamefont{Seiberg}},
  \bibinfo{journal}{Phys. Lett.} \textbf{\bibinfo{volume}{B309}},
  \bibinfo{pages}{337} (\bibinfo{year}{1993}), \eprint{hep-ph/9304307}.

\bibitem[{\citenamefont{Kaplan and Schmaltz}(1994)}]{Kaplan:1993ej}
\bibinfo{author}{\bibfnamefont{D.~B.} \bibnamefont{Kaplan}} \bibnamefont{and}
  \bibinfo{author}{\bibfnamefont{M.}~\bibnamefont{Schmaltz}},
  \bibinfo{journal}{Phys. Rev.} \textbf{\bibinfo{volume}{D49}},
  \bibinfo{pages}{3741} (\bibinfo{year}{1994}), \eprint{hep-ph/9311281}.

\bibitem[{\citenamefont{Arkani-Hamed et~al.}(1996)\citenamefont{Arkani-Hamed,
  Carone, Hall, and Murayama}}]{ArkaniHamed:1996xm}
\bibinfo{author}{\bibfnamefont{N.}~\bibnamefont{Arkani-Hamed}},
  \bibinfo{author}{\bibfnamefont{C.~D.} \bibnamefont{Carone}},
  \bibinfo{author}{\bibfnamefont{L.~J.} \bibnamefont{Hall}}, \bibnamefont{and}
  \bibinfo{author}{\bibfnamefont{H.}~\bibnamefont{Murayama}},
  \bibinfo{journal}{Phys. Rev.} \textbf{\bibinfo{volume}{D54}},
  \bibinfo{pages}{7032} (\bibinfo{year}{1996}), \eprint{hep-ph/9607298}.

\bibitem[{\citenamefont{Kribs et~al.}(2008)\citenamefont{Kribs, Poppitz, and
  Weiner}}]{Kribs:2007ac}
\bibinfo{author}{\bibfnamefont{G.~D.} \bibnamefont{Kribs}},
  \bibinfo{author}{\bibfnamefont{E.}~\bibnamefont{Poppitz}}, \bibnamefont{and}
  \bibinfo{author}{\bibfnamefont{N.}~\bibnamefont{Weiner}},
  \bibinfo{journal}{Phys. Rev.} \textbf{\bibinfo{volume}{D78}},
  \bibinfo{pages}{055010} (\bibinfo{year}{2008}), \eprint{0712.2039}.

\bibitem[{\citenamefont{Dine et~al.}(1996)\citenamefont{Dine, Nelson, Nir, and
  Shirman}}]{Dine:1995ag}
\bibinfo{author}{\bibfnamefont{M.}~\bibnamefont{Dine}},
  \bibinfo{author}{\bibfnamefont{A.~E.} \bibnamefont{Nelson}},
  \bibinfo{author}{\bibfnamefont{Y.}~\bibnamefont{Nir}}, \bibnamefont{and}
  \bibinfo{author}{\bibfnamefont{Y.}~\bibnamefont{Shirman}},
  \bibinfo{journal}{Phys. Rev.} \textbf{\bibinfo{volume}{D53}},
  \bibinfo{pages}{2658} (\bibinfo{year}{1996}), \eprint{hep-ph/9507378}.

\bibitem[{\citenamefont{Carpenter et~al.}(2005)\citenamefont{Carpenter, Fox,
  and Kaplan}}]{Carpenter:2005tz}
\bibinfo{author}{\bibfnamefont{L.~M.} \bibnamefont{Carpenter}},
  \bibinfo{author}{\bibfnamefont{P.~J.} \bibnamefont{Fox}}, \bibnamefont{and}
  \bibinfo{author}{\bibfnamefont{D.~E.} \bibnamefont{Kaplan}}
  (\bibinfo{year}{2005}), \eprint{hep-ph/0503093}.

\bibitem[{\citenamefont{Dumitrescu et~al.}(2010)\citenamefont{Dumitrescu,
  Komargodski, and Sudano}}]{Dumitrescu:2010ca}
\bibinfo{author}{\bibfnamefont{T.~T.} \bibnamefont{Dumitrescu}},
  \bibinfo{author}{\bibfnamefont{Z.}~\bibnamefont{Komargodski}},
  \bibnamefont{and} \bibinfo{author}{\bibfnamefont{M.}~\bibnamefont{Sudano}}
  (\bibinfo{year}{2010}), \eprint{1007.5352}.

\bibitem[{\citenamefont{Nelson and Seiberg}(1994)}]{Nelson:1993nf}
\bibinfo{author}{\bibfnamefont{A.~E.} \bibnamefont{Nelson}} \bibnamefont{and}
  \bibinfo{author}{\bibfnamefont{N.}~\bibnamefont{Seiberg}},
  \bibinfo{journal}{Nucl. Phys.} \textbf{\bibinfo{volume}{B416}},
  \bibinfo{pages}{46} (\bibinfo{year}{1994}), \eprint{hep-ph/9309299}.

\bibitem[{\citenamefont{Arkani-Hamed et~al.}(2005)\citenamefont{Arkani-Hamed,
  Dimopoulos, Giudice, and Romanino}}]{ArkaniHamed:2004yi}
\bibinfo{author}{\bibfnamefont{N.}~\bibnamefont{Arkani-Hamed}},
  \bibinfo{author}{\bibfnamefont{S.}~\bibnamefont{Dimopoulos}},
  \bibinfo{author}{\bibfnamefont{G.~F.} \bibnamefont{Giudice}},
  \bibnamefont{and} \bibinfo{author}{\bibfnamefont{A.}~\bibnamefont{Romanino}},
  \bibinfo{journal}{Nucl. Phys.} \textbf{\bibinfo{volume}{B709}},
  \bibinfo{pages}{3} (\bibinfo{year}{2005}), \eprint{hep-ph/0409232}.

\bibitem[{\citenamefont{Bagger et~al.}(1994)\citenamefont{Bagger, Poppitz, and
  Randall}}]{Bagger:1994hh}
\bibinfo{author}{\bibfnamefont{J.}~\bibnamefont{Bagger}},
  \bibinfo{author}{\bibfnamefont{E.}~\bibnamefont{Poppitz}}, \bibnamefont{and}
  \bibinfo{author}{\bibfnamefont{L.}~\bibnamefont{Randall}},
  \bibinfo{journal}{Nucl. Phys.} \textbf{\bibinfo{volume}{B426}},
  \bibinfo{pages}{3} (\bibinfo{year}{1994}), \eprint{hep-ph/9405345}.

\bibitem[{\citenamefont{Giudice et~al.}(1998)\citenamefont{Giudice, Luty,
  Murayama, and Rattazzi}}]{Giudice:1998xp}
\bibinfo{author}{\bibfnamefont{G.~F.} \bibnamefont{Giudice}},
  \bibinfo{author}{\bibfnamefont{M.~A.} \bibnamefont{Luty}},
  \bibinfo{author}{\bibfnamefont{H.}~\bibnamefont{Murayama}}, \bibnamefont{and}
  \bibinfo{author}{\bibfnamefont{R.}~\bibnamefont{Rattazzi}},
  \bibinfo{journal}{JHEP} \textbf{\bibinfo{volume}{12}}, \bibinfo{pages}{027}
  (\bibinfo{year}{1998}), \eprint{hep-ph/9810442}.

\bibitem[{\citenamefont{Randall and Sundrum}(1999)}]{Randall:1998uk}
\bibinfo{author}{\bibfnamefont{L.}~\bibnamefont{Randall}} \bibnamefont{and}
  \bibinfo{author}{\bibfnamefont{R.}~\bibnamefont{Sundrum}},
  \bibinfo{journal}{Nucl. Phys.} \textbf{\bibinfo{volume}{B557}},
  \bibinfo{pages}{79} (\bibinfo{year}{1999}), \eprint{hep-th/9810155}.

\bibitem[{\citenamefont{Dine et~al.}(1981)\citenamefont{Dine, Fischler, and
  Srednicki}}]{Dine:1981za}
\bibinfo{author}{\bibfnamefont{M.}~\bibnamefont{Dine}},
  \bibinfo{author}{\bibfnamefont{W.}~\bibnamefont{Fischler}}, \bibnamefont{and}
  \bibinfo{author}{\bibfnamefont{M.}~\bibnamefont{Srednicki}},
  \bibinfo{journal}{Nucl. Phys.} \textbf{\bibinfo{volume}{B189}},
  \bibinfo{pages}{575} (\bibinfo{year}{1981}).

\bibitem[{\citenamefont{Dine and Fischler}(1982)}]{Dine:1981gu}
\bibinfo{author}{\bibfnamefont{M.}~\bibnamefont{Dine}} \bibnamefont{and}
  \bibinfo{author}{\bibfnamefont{W.}~\bibnamefont{Fischler}},
  \bibinfo{journal}{Phys. Lett.} \textbf{\bibinfo{volume}{B110}},
  \bibinfo{pages}{227} (\bibinfo{year}{1982}).

\bibitem[{\citenamefont{Dimopoulos
  et~al.}(1981{\natexlab{b}})\citenamefont{Dimopoulos, Raby, and
  Kane}}]{Dimopoulos:1980yf}
\bibinfo{author}{\bibfnamefont{S.}~\bibnamefont{Dimopoulos}},
  \bibinfo{author}{\bibfnamefont{S.}~\bibnamefont{Raby}}, \bibnamefont{and}
  \bibinfo{author}{\bibfnamefont{G.~L.} \bibnamefont{Kane}},
  \bibinfo{journal}{Nucl. Phys.} \textbf{\bibinfo{volume}{B182}},
  \bibinfo{pages}{77} (\bibinfo{year}{1981}{\natexlab{b}}).

\bibitem[{\citenamefont{Nappi and Ovrut}(1982)}]{Nappi:1982hm}
\bibinfo{author}{\bibfnamefont{C.~R.} \bibnamefont{Nappi}} \bibnamefont{and}
  \bibinfo{author}{\bibfnamefont{B.~A.} \bibnamefont{Ovrut}},
  \bibinfo{journal}{Phys. Lett.} \textbf{\bibinfo{volume}{B113}},
  \bibinfo{pages}{175} (\bibinfo{year}{1982}).

\bibitem[{\citenamefont{Alvarez-Gaume et~al.}(1982)\citenamefont{Alvarez-Gaume,
  Claudson, and Wise}}]{AlvarezGaume:1981wy}
\bibinfo{author}{\bibfnamefont{L.}~\bibnamefont{Alvarez-Gaume}},
  \bibinfo{author}{\bibfnamefont{M.}~\bibnamefont{Claudson}}, \bibnamefont{and}
  \bibinfo{author}{\bibfnamefont{M.~B.} \bibnamefont{Wise}},
  \bibinfo{journal}{Nucl. Phys.} \textbf{\bibinfo{volume}{B207}},
  \bibinfo{pages}{96} (\bibinfo{year}{1982}).

\bibitem[{\citenamefont{Dimopoulos and Raby}(1983)}]{Dimopoulos:1982gm}
\bibinfo{author}{\bibfnamefont{S.}~\bibnamefont{Dimopoulos}} \bibnamefont{and}
  \bibinfo{author}{\bibfnamefont{S.}~\bibnamefont{Raby}},
  \bibinfo{journal}{Nucl. Phys.} \textbf{\bibinfo{volume}{B219}},
  \bibinfo{pages}{479} (\bibinfo{year}{1983}).

\bibitem[{\citenamefont{Dine and Nelson}(1993)}]{Dine:1993yw}
\bibinfo{author}{\bibfnamefont{M.}~\bibnamefont{Dine}} \bibnamefont{and}
  \bibinfo{author}{\bibfnamefont{A.~E.} \bibnamefont{Nelson}},
  \bibinfo{journal}{Phys. Rev.} \textbf{\bibinfo{volume}{D48}},
  \bibinfo{pages}{1277} (\bibinfo{year}{1993}), \eprint{hep-ph/9303230}.

\bibitem[{\citenamefont{Dine et~al.}(1995)\citenamefont{Dine, Nelson, and
  Shirman}}]{Dine:1994vc}
\bibinfo{author}{\bibfnamefont{M.}~\bibnamefont{Dine}},
  \bibinfo{author}{\bibfnamefont{A.~E.} \bibnamefont{Nelson}},
  \bibnamefont{and} \bibinfo{author}{\bibfnamefont{Y.}~\bibnamefont{Shirman}},
  \bibinfo{journal}{Phys. Rev.} \textbf{\bibinfo{volume}{D51}},
  \bibinfo{pages}{1362} (\bibinfo{year}{1995}), \eprint{hep-ph/9408384}.

\bibitem[{\citenamefont{Kaplan et~al.}(2000)\citenamefont{Kaplan, Kribs, and
  Schmaltz}}]{Kaplan:1999ac}
\bibinfo{author}{\bibfnamefont{D.~E.} \bibnamefont{Kaplan}},
  \bibinfo{author}{\bibfnamefont{G.~D.} \bibnamefont{Kribs}}, \bibnamefont{and}
  \bibinfo{author}{\bibfnamefont{M.}~\bibnamefont{Schmaltz}},
  \bibinfo{journal}{Phys. Rev.} \textbf{\bibinfo{volume}{D62}},
  \bibinfo{pages}{035010} (\bibinfo{year}{2000}), \eprint{hep-ph/9911293}.

\bibitem[{\citenamefont{Chacko et~al.}(2000)\citenamefont{Chacko, Luty, Nelson,
  and Ponton}}]{Chacko:1999mi}
\bibinfo{author}{\bibfnamefont{Z.}~\bibnamefont{Chacko}},
  \bibinfo{author}{\bibfnamefont{M.~A.} \bibnamefont{Luty}},
  \bibinfo{author}{\bibfnamefont{A.~E.} \bibnamefont{Nelson}},
  \bibnamefont{and} \bibinfo{author}{\bibfnamefont{E.}~\bibnamefont{Ponton}},
  \bibinfo{journal}{JHEP} \textbf{\bibinfo{volume}{01}}, \bibinfo{pages}{003}
  (\bibinfo{year}{2000}), \eprint{hep-ph/9911323}.

\bibitem[{\citenamefont{Polchinski and Susskind}(1982)}]{Polchinski:1982an}
\bibinfo{author}{\bibfnamefont{J.}~\bibnamefont{Polchinski}} \bibnamefont{and}
  \bibinfo{author}{\bibfnamefont{L.}~\bibnamefont{Susskind}},
  \bibinfo{journal}{Phys. Rev.} \textbf{\bibinfo{volume}{D26}},
  \bibinfo{pages}{3661} (\bibinfo{year}{1982}).

\bibitem[{\citenamefont{Hall and Randall}(1991)}]{Hall:1990hq}
\bibinfo{author}{\bibfnamefont{L.~J.} \bibnamefont{Hall}} \bibnamefont{and}
  \bibinfo{author}{\bibfnamefont{L.}~\bibnamefont{Randall}},
  \bibinfo{journal}{Nucl. Phys.} \textbf{\bibinfo{volume}{B352}},
  \bibinfo{pages}{289} (\bibinfo{year}{1991}).

\bibitem[{\citenamefont{Randall and Rius}(1992)}]{Randall:1992cq}
\bibinfo{author}{\bibfnamefont{L.}~\bibnamefont{Randall}} \bibnamefont{and}
  \bibinfo{author}{\bibfnamefont{N.}~\bibnamefont{Rius}},
  \bibinfo{journal}{Phys. Lett.} \textbf{\bibinfo{volume}{B286}},
  \bibinfo{pages}{299} (\bibinfo{year}{1992}).

\bibitem[{\citenamefont{Dine and MacIntire}(1992)}]{Dine:1992yw}
\bibinfo{author}{\bibfnamefont{M.}~\bibnamefont{Dine}} \bibnamefont{and}
  \bibinfo{author}{\bibfnamefont{D.}~\bibnamefont{MacIntire}},
  \bibinfo{journal}{Phys. Rev.} \textbf{\bibinfo{volume}{D46}},
  \bibinfo{pages}{2594} (\bibinfo{year}{1992}), \eprint{hep-ph/9205227}.

\bibitem[{\citenamefont{Fox et~al.}(2002)\citenamefont{Fox, Nelson, and
  Weiner}}]{Fox:2002bu}
\bibinfo{author}{\bibfnamefont{P.~J.} \bibnamefont{Fox}},
  \bibinfo{author}{\bibfnamefont{A.~E.} \bibnamefont{Nelson}},
  \bibnamefont{and} \bibinfo{author}{\bibfnamefont{N.}~\bibnamefont{Weiner}},
  \bibinfo{journal}{JHEP} \textbf{\bibinfo{volume}{08}}, \bibinfo{pages}{035}
  (\bibinfo{year}{2002}), \eprint{hep-ph/0206096}.

\bibitem[{\citenamefont{Cohen et~al.}(2007)\citenamefont{Cohen, Roy, and
  Schmaltz}}]{Cohen:2006qc}
\bibinfo{author}{\bibfnamefont{A.~G.} \bibnamefont{Cohen}},
  \bibinfo{author}{\bibfnamefont{T.~S.} \bibnamefont{Roy}}, \bibnamefont{and}
  \bibinfo{author}{\bibfnamefont{M.}~\bibnamefont{Schmaltz}},
  \bibinfo{journal}{JHEP} \textbf{\bibinfo{volume}{02}}, \bibinfo{pages}{027}
  (\bibinfo{year}{2007}), \eprint{hep-ph/0612100}.

\bibitem[{\citenamefont{Luty and Sundrum}(2002)}]{Luty:2001jh}
\bibinfo{author}{\bibfnamefont{M.~A.} \bibnamefont{Luty}} \bibnamefont{and}
  \bibinfo{author}{\bibfnamefont{R.}~\bibnamefont{Sundrum}},
  \bibinfo{journal}{Phys. Rev.} \textbf{\bibinfo{volume}{D65}},
  \bibinfo{pages}{066004} (\bibinfo{year}{2002}), \eprint{hep-th/0105137}.

\bibitem[{\citenamefont{Luty and Sundrum}(2003)}]{Luty:2001zv}
\bibinfo{author}{\bibfnamefont{M.}~\bibnamefont{Luty}} \bibnamefont{and}
  \bibinfo{author}{\bibfnamefont{R.}~\bibnamefont{Sundrum}},
  \bibinfo{journal}{Phys. Rev.} \textbf{\bibinfo{volume}{D67}},
  \bibinfo{pages}{045007} (\bibinfo{year}{2003}), \eprint{hep-th/0111231}.

\bibitem[{\citenamefont{Dine et~al.}(2004)}]{Dine:2004dv}
\bibinfo{author}{\bibfnamefont{M.}~\bibnamefont{Dine}} \bibnamefont{et~al.},
  \bibinfo{journal}{Phys. Rev.} \textbf{\bibinfo{volume}{D70}},
  \bibinfo{pages}{045023} (\bibinfo{year}{2004}), \eprint{hep-ph/0405159}.

\bibitem[{\citenamefont{Ibe et~al.}(2006{\natexlab{a}})\citenamefont{Ibe,
  Izawa, Nakayama, Shinbara, and Yanagida}}]{Ibe:2005pj}
\bibinfo{author}{\bibfnamefont{M.}~\bibnamefont{Ibe}},
  \bibinfo{author}{\bibfnamefont{K.~I.} \bibnamefont{Izawa}},
  \bibinfo{author}{\bibfnamefont{Y.}~\bibnamefont{Nakayama}},
  \bibinfo{author}{\bibfnamefont{Y.}~\bibnamefont{Shinbara}}, \bibnamefont{and}
  \bibinfo{author}{\bibfnamefont{T.}~\bibnamefont{Yanagida}},
  \bibinfo{journal}{Phys. Rev.} \textbf{\bibinfo{volume}{D73}},
  \bibinfo{pages}{015004} (\bibinfo{year}{2006}{\natexlab{a}}),
  \eprint{hep-ph/0506023}.

\bibitem[{\citenamefont{Ibe et~al.}(2006{\natexlab{b}})\citenamefont{Ibe,
  Izawa, Nakayama, Shinbara, and Yanagida}}]{Ibe:2005qv}
\bibinfo{author}{\bibfnamefont{M.}~\bibnamefont{Ibe}},
  \bibinfo{author}{\bibfnamefont{K.~I.} \bibnamefont{Izawa}},
  \bibinfo{author}{\bibfnamefont{Y.}~\bibnamefont{Nakayama}},
  \bibinfo{author}{\bibfnamefont{Y.}~\bibnamefont{Shinbara}}, \bibnamefont{and}
  \bibinfo{author}{\bibfnamefont{T.}~\bibnamefont{Yanagida}},
  \bibinfo{journal}{Phys. Rev.} \textbf{\bibinfo{volume}{D73}},
  \bibinfo{pages}{035012} (\bibinfo{year}{2006}{\natexlab{b}}),
  \eprint{hep-ph/0509229}.

\bibitem[{\citenamefont{Schmaltz and Sundrum}(2006)}]{Schmaltz:2006qs}
\bibinfo{author}{\bibfnamefont{M.}~\bibnamefont{Schmaltz}} \bibnamefont{and}
  \bibinfo{author}{\bibfnamefont{R.}~\bibnamefont{Sundrum}},
  \bibinfo{journal}{JHEP} \textbf{\bibinfo{volume}{11}}, \bibinfo{pages}{011}
  (\bibinfo{year}{2006}), \eprint{hep-th/0608051}.

\bibitem[{\citenamefont{Nelson et~al.}(2002)\citenamefont{Nelson, Rius, Sanz,
  and Unsal}}]{Nelson:2002ca}
\bibinfo{author}{\bibfnamefont{A.~E.} \bibnamefont{Nelson}},
  \bibinfo{author}{\bibfnamefont{N.}~\bibnamefont{Rius}},
  \bibinfo{author}{\bibfnamefont{V.}~\bibnamefont{Sanz}}, \bibnamefont{and}
  \bibinfo{author}{\bibfnamefont{M.}~\bibnamefont{Unsal}},
  \bibinfo{journal}{JHEP} \textbf{\bibinfo{volume}{08}}, \bibinfo{pages}{039}
  (\bibinfo{year}{2002}), \eprint{hep-ph/0206102}.

\bibitem[{\citenamefont{Chacko et~al.}(2005)\citenamefont{Chacko, Fox, and
  Murayama}}]{Chacko:2004mi}
\bibinfo{author}{\bibfnamefont{Z.}~\bibnamefont{Chacko}},
  \bibinfo{author}{\bibfnamefont{P.~J.} \bibnamefont{Fox}}, \bibnamefont{and}
  \bibinfo{author}{\bibfnamefont{H.}~\bibnamefont{Murayama}},
  \bibinfo{journal}{Nucl. Phys.} \textbf{\bibinfo{volume}{B706}},
  \bibinfo{pages}{53} (\bibinfo{year}{2005}), \eprint{hep-ph/0406142}.

\bibitem[{\citenamefont{Carone et~al.}(2005)\citenamefont{Carone, Erlich, and
  Glover}}]{Carone:2005iq}
\bibinfo{author}{\bibfnamefont{C.~D.} \bibnamefont{Carone}},
  \bibinfo{author}{\bibfnamefont{J.}~\bibnamefont{Erlich}}, \bibnamefont{and}
  \bibinfo{author}{\bibfnamefont{B.}~\bibnamefont{Glover}},
  \bibinfo{journal}{JHEP} \textbf{\bibinfo{volume}{10}}, \bibinfo{pages}{042}
  (\bibinfo{year}{2005}), \eprint{hep-ph/0509002}.

\bibitem[{\citenamefont{Nomura et~al.}(2006)\citenamefont{Nomura, Poland, and
  Tweedie}}]{Nomura:2005rj}
\bibinfo{author}{\bibfnamefont{Y.}~\bibnamefont{Nomura}},
  \bibinfo{author}{\bibfnamefont{D.}~\bibnamefont{Poland}}, \bibnamefont{and}
  \bibinfo{author}{\bibfnamefont{B.}~\bibnamefont{Tweedie}},
  \bibinfo{journal}{Nucl. Phys.} \textbf{\bibinfo{volume}{B745}},
  \bibinfo{pages}{29} (\bibinfo{year}{2006}), \eprint{hep-ph/0509243}.

\bibitem[{\citenamefont{Nomura and Tweedie}(2005)}]{Nomura:2005qg}
\bibinfo{author}{\bibfnamefont{Y.}~\bibnamefont{Nomura}} \bibnamefont{and}
  \bibinfo{author}{\bibfnamefont{B.}~\bibnamefont{Tweedie}},
  \bibinfo{journal}{Phys. Rev.} \textbf{\bibinfo{volume}{D72}},
  \bibinfo{pages}{015006} (\bibinfo{year}{2005}), \eprint{hep-ph/0504246}.

\bibitem[{\citenamefont{Antoniadis et~al.}(2008)\citenamefont{Antoniadis,
  Benakli, Delgado, and Quiros}}]{Antoniadis:2006uj}
\bibinfo{author}{\bibfnamefont{I.}~\bibnamefont{Antoniadis}},
  \bibinfo{author}{\bibfnamefont{K.}~\bibnamefont{Benakli}},
  \bibinfo{author}{\bibfnamefont{A.}~\bibnamefont{Delgado}}, \bibnamefont{and}
  \bibinfo{author}{\bibfnamefont{M.}~\bibnamefont{Quiros}},
  \bibinfo{journal}{Adv. Stud. Theor. Phys.} \textbf{\bibinfo{volume}{2}},
  \bibinfo{pages}{645} (\bibinfo{year}{2008}), \eprint{hep-ph/0610265}.

\bibitem[{\citenamefont{Nakayama et~al.}(2007)\citenamefont{Nakayama, Taki,
  Watari, and Yanagida}}]{Nakayama:2007cf}
\bibinfo{author}{\bibfnamefont{Y.}~\bibnamefont{Nakayama}},
  \bibinfo{author}{\bibfnamefont{M.}~\bibnamefont{Taki}},
  \bibinfo{author}{\bibfnamefont{T.}~\bibnamefont{Watari}}, \bibnamefont{and}
  \bibinfo{author}{\bibfnamefont{T.~T.} \bibnamefont{Yanagida}},
  \bibinfo{journal}{Phys. Lett.} \textbf{\bibinfo{volume}{B655}},
  \bibinfo{pages}{58} (\bibinfo{year}{2007}), \eprint{0705.0865}.

\bibitem[{\citenamefont{Amigo et~al.}(2009)\citenamefont{Amigo, Blechman, Fox,
  and Poppitz}}]{Amigo:2008rc}
\bibinfo{author}{\bibfnamefont{S.~D.~L.} \bibnamefont{Amigo}},
  \bibinfo{author}{\bibfnamefont{A.~E.} \bibnamefont{Blechman}},
  \bibinfo{author}{\bibfnamefont{P.~J.} \bibnamefont{Fox}}, \bibnamefont{and}
  \bibinfo{author}{\bibfnamefont{E.}~\bibnamefont{Poppitz}},
  \bibinfo{journal}{JHEP} \textbf{\bibinfo{volume}{01}}, \bibinfo{pages}{018}
  (\bibinfo{year}{2009}), \eprint{0809.1112}.

\bibitem[{\citenamefont{Kribs et~al.}(2009{\natexlab{a}})\citenamefont{Kribs,
  Martin, and Roy}}]{Kribs:2009zy}
\bibinfo{author}{\bibfnamefont{G.~D.} \bibnamefont{Kribs}},
  \bibinfo{author}{\bibfnamefont{A.}~\bibnamefont{Martin}}, \bibnamefont{and}
  \bibinfo{author}{\bibfnamefont{T.~S.} \bibnamefont{Roy}},
  \bibinfo{journal}{JHEP} \textbf{\bibinfo{volume}{06}}, \bibinfo{pages}{042}
  (\bibinfo{year}{2009}{\natexlab{a}}), \eprint{0901.4105}.

\bibitem[{\citenamefont{Herquet et~al.}(2010)\citenamefont{Herquet, Knegjens,
  and Laenen}}]{Herquet:2010ka}
\bibinfo{author}{\bibfnamefont{M.}~\bibnamefont{Herquet}},
  \bibinfo{author}{\bibfnamefont{R.}~\bibnamefont{Knegjens}}, \bibnamefont{and}
  \bibinfo{author}{\bibfnamefont{E.}~\bibnamefont{Laenen}}
  (\bibinfo{year}{2010}), \eprint{1005.2900}.

\bibitem[{\citenamefont{Kribs et~al.}(2009{\natexlab{b}})\citenamefont{Kribs,
  Martin, and Roy}}]{Kribs:2008hq}
\bibinfo{author}{\bibfnamefont{G.~D.} \bibnamefont{Kribs}},
  \bibinfo{author}{\bibfnamefont{A.}~\bibnamefont{Martin}}, \bibnamefont{and}
  \bibinfo{author}{\bibfnamefont{T.~S.} \bibnamefont{Roy}},
  \bibinfo{journal}{JHEP} \textbf{\bibinfo{volume}{01}}, \bibinfo{pages}{023}
  (\bibinfo{year}{2009}{\natexlab{b}}), \eprint{0807.4936}.

\bibitem[{\citenamefont{Blechman and Ng}(2008)}]{Blechman:2008gu}
\bibinfo{author}{\bibfnamefont{A.~E.} \bibnamefont{Blechman}} \bibnamefont{and}
  \bibinfo{author}{\bibfnamefont{S.-P.} \bibnamefont{Ng}},
  \bibinfo{journal}{JHEP} \textbf{\bibinfo{volume}{06}}, \bibinfo{pages}{043}
  (\bibinfo{year}{2008}), \eprint{0803.3811}.

\bibitem[{\citenamefont{Fok and Kribs}(2010)}]{Fok:2010vk}
\bibinfo{author}{\bibfnamefont{R.}~\bibnamefont{Fok}} \bibnamefont{and}
  \bibinfo{author}{\bibfnamefont{G.~D.} \bibnamefont{Kribs}}
  (\bibinfo{year}{2010}), \eprint{1004.0556}.

\bibitem[{\citenamefont{Roy and Schmaltz}(2008)}]{Roy:2007nz}
\bibinfo{author}{\bibfnamefont{T.~S.} \bibnamefont{Roy}} \bibnamefont{and}
  \bibinfo{author}{\bibfnamefont{M.}~\bibnamefont{Schmaltz}},
  \bibinfo{journal}{Phys. Rev.} \textbf{\bibinfo{volume}{D77}},
  \bibinfo{pages}{095008} (\bibinfo{year}{2008}), \eprint{0708.3593}.

\bibitem[{\citenamefont{Murayama et~al.}(2008)\citenamefont{Murayama, Nomura,
  and Poland}}]{Murayama:2007ge}
\bibinfo{author}{\bibfnamefont{H.}~\bibnamefont{Murayama}},
  \bibinfo{author}{\bibfnamefont{Y.}~\bibnamefont{Nomura}}, \bibnamefont{and}
  \bibinfo{author}{\bibfnamefont{D.}~\bibnamefont{Poland}},
  \bibinfo{journal}{Phys. Rev.} \textbf{\bibinfo{volume}{D77}},
  \bibinfo{pages}{015005} (\bibinfo{year}{2008}), \eprint{0709.0775}.

\bibitem[{\citenamefont{Perez et~al.}(2009)\citenamefont{Perez, Roy, and
  Schmaltz}}]{Perez:2008ng}
\bibinfo{author}{\bibfnamefont{G.}~\bibnamefont{Perez}},
  \bibinfo{author}{\bibfnamefont{T.~S.} \bibnamefont{Roy}}, \bibnamefont{and}
  \bibinfo{author}{\bibfnamefont{M.}~\bibnamefont{Schmaltz}},
  \bibinfo{journal}{Phys. Rev.} \textbf{\bibinfo{volume}{D79}},
  \bibinfo{pages}{095016} (\bibinfo{year}{2009}), \eprint{0811.3206}.

\bibitem[{\citenamefont{Deser and Zumino}(1977)}]{Deser:1977uq}
\bibinfo{author}{\bibfnamefont{S.}~\bibnamefont{Deser}} \bibnamefont{and}
  \bibinfo{author}{\bibfnamefont{B.}~\bibnamefont{Zumino}},
  \bibinfo{journal}{Phys. Rev. Lett.} \textbf{\bibinfo{volume}{38}},
  \bibinfo{pages}{1433} (\bibinfo{year}{1977}).

\bibitem[{\citenamefont{Cacciapaglia et~al.}(2006)\citenamefont{Cacciapaglia,
  Csaki, Marandella, and Strumia}}]{Cacciapaglia:2006pk}
\bibinfo{author}{\bibfnamefont{G.}~\bibnamefont{Cacciapaglia}},
  \bibinfo{author}{\bibfnamefont{C.}~\bibnamefont{Csaki}},
  \bibinfo{author}{\bibfnamefont{G.}~\bibnamefont{Marandella}},
  \bibnamefont{and} \bibinfo{author}{\bibfnamefont{A.}~\bibnamefont{Strumia}},
  \bibinfo{journal}{Phys. Rev.} \textbf{\bibinfo{volume}{D74}},
  \bibinfo{pages}{033011} (\bibinfo{year}{2006}), \eprint{hep-ph/0604111}.

\bibitem[{\citenamefont{Hook et~al.}(2010)\citenamefont{Hook, Izaguirre, and
  Wacker}}]{Hook:2010tw}
\bibinfo{author}{\bibfnamefont{A.}~\bibnamefont{Hook}},
  \bibinfo{author}{\bibfnamefont{E.}~\bibnamefont{Izaguirre}},
  \bibnamefont{and} \bibinfo{author}{\bibfnamefont{J.~G.} \bibnamefont{Wacker}}
  (\bibinfo{year}{2010}), \eprint{1006.0973}.

\bibitem[{\citenamefont{Bagger et~al.}(1995)\citenamefont{Bagger, Poppitz, and
  Randall}}]{Bagger:1995ay}
\bibinfo{author}{\bibfnamefont{J.}~\bibnamefont{Bagger}},
  \bibinfo{author}{\bibfnamefont{E.}~\bibnamefont{Poppitz}}, \bibnamefont{and}
  \bibinfo{author}{\bibfnamefont{L.}~\bibnamefont{Randall}},
  \bibinfo{journal}{Nucl. Phys.} \textbf{\bibinfo{volume}{B455}},
  \bibinfo{pages}{59} (\bibinfo{year}{1995}), \eprint{hep-ph/9505244}.

\end{thebibliography}

\end{document}